\documentclass[a4paper,fleqn,usenatbib]{mnras}

\usepackage{newtxtext,newtxmath}

\usepackage[T1]{fontenc}
\usepackage{ae,aecompl}

\usepackage{graphicx}	
\usepackage{grffile}
\usepackage{epstopdf}
\usepackage{amsmath}	
\usepackage{amssymb}	

\def\be{\begin{equation}}
\def\ee{\end{equation}}

\def\bea{\begin{eqnarray}}
\def\eea{\end{eqnarray}}

\def\bite{\begin{itemize}}
\def\eite{\end{itemize}}

\title[On Radiative Acceleration in GRB Fireballs]{Dynamic Monte Carlo Simulations of Radiatively Accelerated GRB Fireballs}

\author[Chhotray \& Lazzati]{
Atul Chhotray,$^{1}$\thanks{E-mail: chhotraa@oregonstate.edu (AC)}
Davide Lazzati$^{1}$
\\
$^{1}$Department of Physics, Weniger Hall, Oregon State University, Corvallis 97331, OR, USA
}

\date{Accepted XXX. Received YYY; in original form ZZZ}

\pubyear{2015}

\begin{document}
\label{firstpage}
\pagerange{\pageref{firstpage}--\pageref{lastpage}}
\maketitle

\begin{abstract}
We present a novel Dynamic Monte Carlo code (DynaMo code) which self-consistently simulates the Compton scattering driven dynamic evolution of a plasma. We use the DynaMo code to investigate the time--dependent expansion and acceleration of dissipationless GRB fireballs by varying their initial opacities and baryonic content. We study the opacity and energy density evolution of an initially optically thick, radiation--dominated fireball across its entire phase space - in particular during the $R_{\rm ph} < R_{\rm sat}$ regime. Our results reveal new phases of fireball evolution: a transition phase with a radial extent of several orders of magnitude - the fireball transitions from $\Gamma \propto R$ to $\Gamma \propto R^0$, a post--photopsheric acceleration phase - where fireballs accelerate beyond the photosphere, and a Thomson--dominated acceleration phase -  characterized by slow acceleration of optically thick, matter--dominated fireballs due to Thomson scattering. 
We quantify the new phases by providing analytical expressions of Lorentz factor evolution, which will be useful for deriving jet parameters.\\
\end{abstract}




\section{Introduction}
Gamma--Ray Bursts (GRBs henceforth) are intense bursts of high energy radiation (high energy X-rays and $\gamma$ rays). First detected serendipitously about five decades ago (\citealt{Klebe_firstGRB73}), today GRBs are regularly detected by space-based satellites and are known to be one of the brightest explosions in the universe ($L\sim 10^{51-52}$ ergs $/$ s) (see \citealt{Piranreview94}). These large luminosities along with small variability time-scales and the emission of high energy gamma rays ($\sim 100+$ keV) led to the compactness problem in GRBs. The compactness problem was resolved by invoking ultra-relativistic motion (\citealt{Paczy86}; \citealt{Goodm86}) of the emitting source, which has been confirmed by observations (\citealt{GRBradioafterglow}; \citealt{frailbeaming})\\*
\\
GRBs are thought to be powered by the core-collapse of a massive rotating star (leading to Long GRBs) and the merger of two neutron stars / a neutron star and black hole (resulting in Short GRBs) (see \citealt{KumaZhang15}). Although ultra-relativistic jets are invoked to explain GRBs, the mechanism responsible for the jet production from GRB progenitors is not well understood and is under investigation. The two mechanisms that have been proposed to launch and accelerate jets to relativistic speeds are 1) magnetic fields and 2) radiation. Several prior works have studied the driving role of magnetic fields in collimating and accelerating relativistic outflows in astrophysical environments (see for e.g., \citealt{BlandZnaje77}, \citealt{McKin06MHDjetform}, \citealt{Komis11}, \citealt{SashaNaraMcKin11jetformMAD}). Consequently, they have been been proposed as the jet production mechanism in GRBs (\citealt{DrenkSprui02}; \citealt{LyutiParie03}). In this paper we will explore in detail radiation as an alternative mechanism to accelerate (and possibly launch) relativistic outflows.\\*
\\
Radiative acceleration of outflows is a well known astrophysical phenomenon (e.g., continuum and line--driven stellar winds -- see \citealt{CAK1975};~\citealt{contiwind06}). To produce relativistic outflows associated with GRBs and AGNs, radiative acceleration due to external radiation sources has been investigated. \cite{MadauThomp00} studied the radiative acceleration of cold, optically thin plasmas due to external radiation sources. \cite{ZampiTurol03} studied radiative acceleration of low density ion-electron plasma by incident transient radiation along with radiation-induced internal electric fields. The above mentioned works have primarily focused on the dynamical evolution of a low-density,  optically thin plasma under the influence of radiation sources external to the plasma. In contrast to the earlier studies, one of the first and foremost theoretical models to understand the physics of GRBs was the GRB fireball model, based on a hot, spherically expanding, outflow optically thick to radiation. The fireball model assumes an initially optically thick plasma, with radiation and matter in thermal equilibrium, and the radiation energy density exceeding the rest mass energy density significantly (\citealt{MesLagunaRees93}; \citealt{Piranreview94}). When these conditions are met, radiation pressure due to the trapped photons dominates the outflow (or fireball - we will use these terms interchangeably) evolution leading to an accelerating, ultra-relativistic fireball.\\*
\\
In this article, we will investigate from a micro-physical perspective Comptonization driven acceleration by radiation advected with the outflow. We present our Dynamic Monte Carlo code (DynaMo code) that we have used to self-consistently investigate the scattering induced acceleration, and relativistic expansion of a spherical fireball. This paper is organized as follows: \S~\ref{sec:theory&meth} begins with an introduction to the fireball theory (\S~\ref{sec:sub_fireball_theory}) followed by an overview of the methodology behind the DynaMo code (\S~\ref{sec:sub_thecode}). In \S~\ref{sec:resanddiscuss} we present and discuss our results, followed by a comparison with the fireball model's theoretical results. In \S~\ref{sec:conclu}, we summarize our work and draw our conclusions.  
\section{Theory and Methodology}\label{sec:theory&meth}
\subsection{Fireball Theory}\label{sec:sub_fireball_theory}
\begin{figure}
	\begin{center}
		\includegraphics[width=1.0\linewidth]{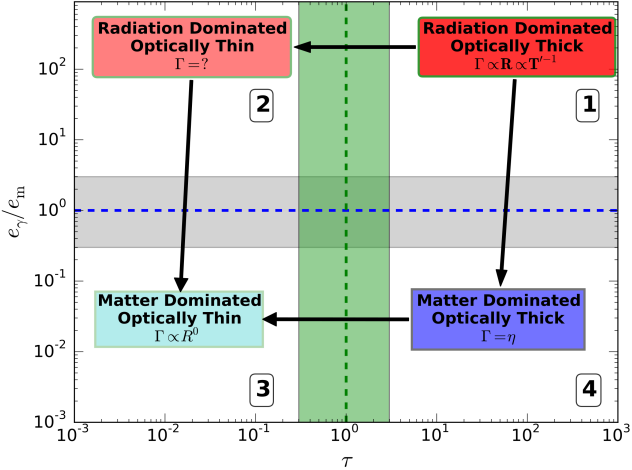}
		\caption{Diagram illustrating the phase space evolution of a GRB fireball based on the fireball model. The y axis depicts the ratio of radiation to matter energy density and the x-axis plots the opacity. In the phase space diagram, GRB fireballs begin their evolution in quadrant 1 ($\tau\gg1, e_{\rm \gamma}/e_{\rm m}\gg1 $) and end it in quadrant 3 ($\tau \ll 1, e_{\rm \gamma}/e_{\rm m}\ll1 $). The arrows show possible evolutionary paths of fireballs. The shaded regions near $\tau \sim 1$ and $e_{\rm \gamma}/e_{\rm m} \sim 1$ represent transition zones or regions where fireball evolution has not been well studied.} 
		\label{fig:fbphasespace}
	\end{center}
\end{figure}
The idea of the fireball was first advocated by \cite{CavalloRees78} to explain the phenomenon of GRBs. Created by depositing a large amount of energy onto matter confined to a small volume, a sufficiently dense fireball will be optically thick ($\tau\gg 1$) to its own radiation. This radiation can drive the acceleration and ultra-relativistic expansion of the fireball if the radiation energy density exceeds the fireball's rest mass energy density. Thus, the fireball model resolved the GRB compactness problem by employing radiation as a mechanism to drive the radiation-matter mixture to ultra-relativistic speeds (\citealt{Paczy86}; \citealt{Goodm86}).
Radiation to matter energy density and opacity are important factors that govern the evolution of a fireball, and this is graphically represented in Fig.~\ref{fig:fbphasespace} by the phase-space diagram. In this diagram, GRB fireballs start optically thick and radiation--dominated in the top--right part of the graph. We refer to this part of the diagram as quadrant 1 (here exponents of both logarithmic axes are positive, analogous to first quadrant in the Cartesian coordinate system where x and y are positive). The fireballs eventually evolve to quadrant 3 (bottom--left portion of Fig.~\ref{fig:fbphasespace} where exponents of both logarithmic axes are negative, similar to the third quadrant in the Cartesian coordinate system) where they become optically thin and matter--dominated. As we discuss later in this section, the evolutionary trajectory of a fireball as it evolves from quadrant 1 to 3 depends on the fireball's initial parameters.\\
To parameterize the fireball the most important physical quantities are 1) the total energy $E$ (includes energies of both matter and radiation), 2) its rest mass $M$,  and 3) the initial radius of the fireball $R_0$ (i.e., fireball evolution begins with bulk Lorentz factor $\Gamma_0=1$ at $R_0$). Radiation-matter interactions transform the fireball's internal energy into bulk kinematic motion.  Assuming that the fireball remains highly opaque ($\tau\gg1$) during expansion, the maximum possible Lorentz factor $\eta$ attained (using energy conservation) is given by -
\be
\eta = \frac{E}{M c^2}.\label{eq:etadef}
\ee
To characterize the evolution (acceleration and expansion) of the fireball plasma, we require the bulk Lorentz factor $\Gamma$ and the radius $R$ of the fireball. These physical quantities do not evolve independently during the radiation--dominated acceleration phase (quadrant 1 in Fig.~\ref{fig:fbphasespace}). The mathematical relationship between these quantities and the comoving temperature $T'$ can be obtained using energy, momentum, and entropy conservation, as given by (\citealt{MesLagunaRees93}; \citealt{Piranreview94}) -- 
\be
\Gamma \propto R \propto \frac{1}{T'} \label{eq:fireball_acc}.
\ee
\cite{Piran1999} has studied an adiabatically expanding fireball with infinite opacity using hydrodynamic equations. The conservation equations for energy, momentum, and number of particles, are as follows:
\bea
R^2 \rho \Gamma & = & c' \\
R^2 e^{\frac{3}{4}}\Gamma & = & c'' \\ 
R^2 \left(\rho + \frac{4e}{3}\right) \Gamma^2 & = & c''', 
\eea
where $R$ denotes the fireball radius, and $c', c''$ and $c'''$ are constants (see \citealt{Piran1999} for an exact definition of the variables used). The boundary conditions for this system are 1) the fireball starts from rest at an initial radius $R_0$ and 2) as $R \rightarrow \infty$ the Lorentz factor equals $\eta$. Using these boundary conditions, the conservation equations can be solved to obtain -
\be
R = \frac{R_0 \Gamma (\eta - 1)^{3/2}}{(\eta - \Gamma)^{3/2}}. \label{eq:altmodelGamma}
\ee
The radiation--dominated acceleration phase can end if 1) the optically thick fireball's radiation energy density becomes comparable to or less than the rest mass energy density (i.e., radiation energy does not dominate matter energy density), and/or 2) radiation escapes at the photopshere due to the expanding plasma's decreasing opacity. In the former case, the trapped photons lose their energy by continually accelerating the plasma to the maximum possible Lorentz factor $\eta$ (see eq.~\ref{eq:etadef}). The fireball, thereafter, coasts at $\eta$ and this new phase is thus termed the saturation or the coasting phase. In the GRB phase space diagram (Fig.~\ref{fig:fbphasespace}), this evolution is represented by the fireball moving from quadrant 1 to 3 via quadrant 4. Using eq.~\ref{eq:fireball_acc}, the characteristic radius at which the saturation phase begins can be computed as -
\be
R_{\rm sat} = \frac{\eta}{\Gamma_0} R_0 = \eta R_0.\label{eq:rsat}
\ee
In the latter case, the photons escape the fireball if the expanding plasma's optical depth does not remain large enough, thereby, bringing an end to the acceleration process. In the phase space diagram, this evolution is represented by the fireball moving from quadrant 1 to 2 and eventually to 3. The characteristic theoretical radius where radiation decouples and escapes the plasma (assumed to occur when the optical depth $\tau \sim 1$) is termed the photospheric radius and is denoted by $R_{\rm ph}$. The absence of radiation causes the driving force to vanish, and as a result the plasma coasts at the Lorentz factor achieved at the photospheric radius.\\
In the previous paragraphs we outlined different evolutionary paths for GRB fireballs. The fireball's evolution depends on its initial energy density and opacity, and the effect of both these parameters can be mathematically represented by the ratio of the characteristic radii, i.e., $R_{\rm ph}/R_{\rm sat}$. It is important to point out that most earlier works have studied extreme regimes of fireball evolution, e.g., the hydrodynamical evolution of a fluid with infinite opacity (\citealt{Piran1999}). \cite{Paczynski1990} explored the evolution of super--eddington winds assuming throughout, 1) an optically thick flow (similar to fireball evolution from quadrant 1 to 4), and 2) a constant velocity past the photosphere. Another well studied scenario is when a radiation--dominated fireball reaches the photosphere and then suddenly loses its  radiation (\citealt{MesLagunaRees93}; \citealt{ReesMesz2005dissphotos}). Earlier works have not studied the realistic scenario where, due to the expansion and acceleration, the fireball 1) gradually becomes optically thin (evolves from quadrant 1 to 2) and loses photons, and 2) is no longer radiation energy dominated (transitions from quadrant 1 to 4). These transition zones are represented by the highlighted regions in Fig.~\ref{fig:fbphasespace}. In addition, analytical calculations encounter difficulties in studying fireball evolution for small opacities (e.g., the Lorentz factor evolution is unknown in quadrant 2 -- as indicated by the question mark in Fig.~\ref{fig:fbphasespace}). In \S~\ref{sec:resanddiscuss}, we will show and discuss the results of fireball evolution in all regimes and across the transition zones. These results are obtained using our DynaMo code, which we detail in the next section.
\subsection{The Code}\label{sec:sub_thecode}
\begin{figure}
	\begin{center}
		\includegraphics[width=0.5\textwidth]{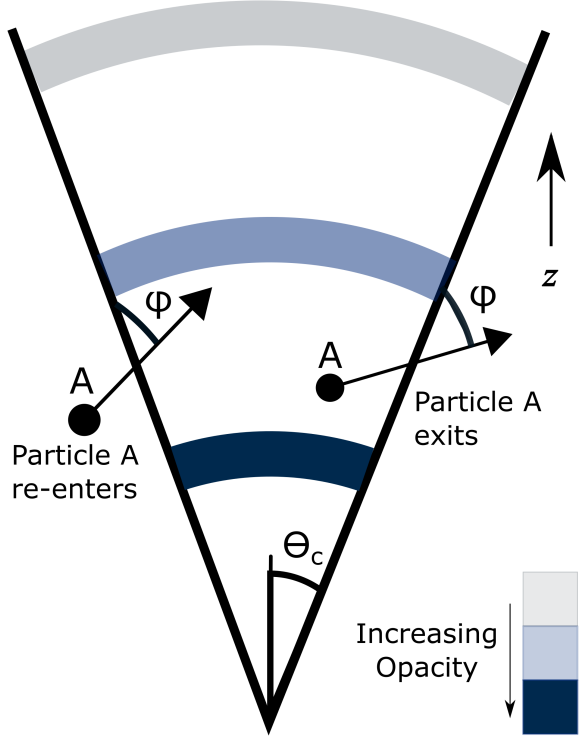}
		\caption{Diagram depicting the cross-sectional view of the simulated wedge. The colored arcs represent a shell traveling radially outward (along the $z$ direction), expanding and becoming more transparent. Also shown is an implementation of the periodic boundary condition with particle A exiting and re-entering the wedge at an angle $\varphi$ with the lateral surface.}
		\label{fig:diagram}
	\end{center}
\end{figure}

In this section we detail the methodology behind the DynaMo code, which simulates the Compton scattering driven expansion and evolution of an outflow. The outflow is a scattering dominated plasma composed of photons, leptons (and protons). Figure~\ref{fig:diagram} shows a diagrammatic cross-sectional view of the outflow geometry, which is a conical wedge with an opening angle $\theta_c$ and encapsulates a section of the fireball's spherical shell. 
Also shown is a fireball shell as it travels radially outward, expands and becomes optically thin (the differently colored arcs represent the expanding shell at different radial positions and opacities).

For our simulations, we are primarily concerned with the physical quantities in three distinct reference frames - 
\bite
\item \textbf{Lab Frame} - The lab or the laboratory frame is the frame at rest with respect to the GRB progenitor producing the outflow or the host galaxy. Any observer at rest in the lab frame observes the jet moving with a bulk Lorentz factor $\Gamma$.

\item \textbf{Jet Frame} - Also known as the comoving frame, this reference frame travels with the GRB outflow at the bulk Lorentz factor $\Gamma$. From the perspective of an observer in the lab frame, the jet frame would be in motion along the radial direction (which is along the z direction as depicted in Fig.~\ref{fig:diagram}). Any physical quantity computed in the jet frame will be primed, e.g., the four--momentum of an electron in the comoving frame will be denoted as $p'^{\mu}_{e^-}$. This frame is naturally suited for calculating local properties of the fluid/plasma (e.g., comoving temperature $T^{\prime}$).

\item \textbf{Particle Frame} - At any given time, the motion of any particle in the plasma will be different from the bulk motion of the jet plasma due to the random thermal motion of the particle. The particle frame is the frame comoving with the lepton (or proton) selected for the scattering (or collision) process (we will discuss the particle interactions in greater depth later in this section). The interaction cross-sections (total and differential) are expressed in terms of physical quantities defined in the frame of the particle involved in the interaction.\\
All physical quantities in the particle frame will be double primed, for example, the four--momentum of a photon as observed in the frame of an electron will be denoted by $p''^{\mu}_{\gamma}$.
\eite

\subsubsection{Fireball Initialization}\label{sec:jet_init}

The first step of our simulation is the initialization of the GRB fireball with the appropriate parameters. As we simulate a section of a fireball shell encapsulated within a conical wedge, the following parameters need to be specified to initialize the simulation -- 
\bite
\item Wedge Opening Angle: $\theta_c = \pi/10^4$ radians. \\
To prevent causality violation and unphysical scatterings, it is imperative that the condition $\theta_c \ll 1/\Gamma$ is always satisfied during fireball evolution.
\item Initial Bulk Lorentz factor: $\Gamma_0 = 1$\\
The GRB fireball initially possesses only internal energy and no bulk kinetic energy.
\item Wedge Outer Radius: $r_{\rm outer,0} = 10^8 $ cm (\citealt{Peer2015Hydroprop})
\item Wedge Inner Radius: $r_{\rm inner,0} = 8 \times 10^7$ cm
\item Particle Initialization:
\bite	
\item Particle Count: The code can simulate interactions between photons, leptons (electrons) and baryons (protons), and thus requires the number $N$ and type of particles to be specified. Unless otherwise stated, the total photon count is $N_{\gamma} = 2800$, and the electron and proton counts are $N_e = N_p = 100$.
\item Spatial Distribution: The code synthesizes the specified number $N$ of particles, and uniformly distributes them within the specified inner and outer radii of the wedge. 
\item Distribution Functions: The user can synthesize particles by supplying the appropriate parameters to the inbuilt energy distribution functions (e.g., comoving temperature $T^{\prime}$ for thermal distribution, power law index $p$ for non-thermal distribution, the energy for mono-energetic distribution).\\
Though the code can accept any temperature value, the simulations shown in this paper start at $T^{\prime} = 7.7 \times 10^9$ K. At this temperature the electrons are distributed according to a Maxwell J{\"{u}}ttner distribution, the protons adopt a Maxwell Boltzmann distribution, and the photons conform to a Blackbody distribution.
\eite
\eite
\subsubsection{Dynamic Time Step Calculation}\label{sec:timestep}
As one of the motivations of our code is to track and evolve the plasma and its constituents, it is essential to track the four--momenta and positions of all the particles in the lab frame. In order to dynamically evolve the system a time step is required, and the maximum value of that step is dictated by the mean free times of the photons in the plasma. Let us first consider photon-lepton scatterings. From a physical perspective, all the particles travel in straight lines until a photon-lepton pair scatters, exchanges energy and momenta, and then the particles again travel in straight lines until the next photon-lepton scattering occurs. This time interval between successive photon-lepton scatterings is the collision-free time experienced by all particles in the plasma (we shall refer to this interval as the collision time - $t_{\rm coll}$). Thus, the dynamic time step in our simulation is this collision time. The reader should refer to Appendix~\ref{sec:Appdx_timestep} for a detailed calculation of the collision time.\\
As experienced by a single photon, the infinitesimal opacity of a medium (at rest) as given in \cite{Rybicki1979} is --
\be
d\tau = n \sigma dl\label{eq:opacity_basic},
\ee
where $n$ denotes the number density of the scatterers, $\sigma$ is the effective Thomson cross section (we explain this in greater detail in Appendix~\ref{sec:calcRphRsat}), and $dl$ is the infinitesimal path length traversed by the photon.\\
Using eq.~\ref{eq:opacity_basic} and as shown in detail in Appendix~\ref{sec:Appdx_timestep}, the mean free time $t_{\rm mf}$ of the photon is (see eq.~\ref{eq:mft_basic}) --
\be
t_{\rm mf} = \frac{1}{n c\sigma},
\ee
where $c = 3 \times 10^{10}$ cm$/$s, is the speed of light. If the scattering medium contains $\rm N_p$ photons, the mean free time $\tau_{\rm Pop}$ for any photon within this population would be given by --
\be
t_{\rm Pop} = \frac{1}{\sum_{i=1}^{\rm N_p} n c \sigma } = \frac{1}{\sum_{i=1}^{\rm N_p} \left(\frac{1}{t_{\rm mf}}\right)} = \frac{t_{\rm mf}}{\rm N_p}.\label{eq:mft_1pho_manylep}
\ee
Physically, if the number of photons in a medium increases so should the likelihood of a photon colliding, thereby decreasing the mean free time of the entire population.\\
Now we shall discuss how the opacity and mean free quantities are modified if the scatterers themselves are in motion. The motion of scatterers introduces a velocity dependence in the opacity. As shown in Appendix~\ref{sec:tstep_appdx_scatmot} and \cite{AbramNovikPaczy91}, a single photon immersed in a medium of moving $\rm N_s$ scatterers, experiences an opacity $d\tau$ given by --
\be
d\tau = \sum_{j=1}^{\rm N_s} n_j \sigma (1 - \beta_j \cos \theta_j) dl,\label{eq:opacity_1pho_gen}
\ee
where $\beta_j$ is the speed of the scatterer traveling in the $\rm j^{\rm th}$ direction normalized by the speed of light and $\theta_j$ is the angle between the three momenta of the photon and the scatterer.\\
Let us now consider a plasma having multiple photons interacting with multiple scatterers. As shown in Appendix~\ref{sec:tstep_appdx_scatmot}, the mean free time of the entire population is --
\be
t_{\rm Pop} = \frac{V}{\sum_i \sum_j \sigma (1 - \beta_{\rm j} \cos \theta_{\rm ij}) c} = \frac{1}{\left(\sum_{i} \left(\sum_{j}\frac{1}{t_{\rm ij}}\right)\right)} = \frac{1}{\sum_{i} \left(\frac{1}{t_{\rm i}}\right)} \label{eq:mft_pop}.
\ee
Thus all the particles experience an \textit{average or mean} free time $t_{\rm Pop}$ given by eq.~\ref{eq:mft_pop}. We note that this is an average value and therefore the extraction of the real collision time or the actual free travel time between successive scatterings (which is different from the mean free time) requires the use of a Monte Carlo acceptance-rejection algorithm. The probability of one photon scattering between the time interval $t$ and $t + dt$ is proportional to --
\be
P(t) \propto e^{\frac{-t}{t_{\rm Pop}}},\label{eq:scatprob}
\ee
which we use to obtain the collision time $t_{\rm coll}$ for our simulation.

\subsubsection{Particle Selection}\label{sec:Part_Sel}
The collision time $t_{\rm coll}$ determines the travel time between photon--lepton scatterings. As some photon--lepton pairs are more likely to scatter than others, the code thus needs to account for this likelihood to identify and select the scattering pair.\\
Intuitively, the photon more likely to scatter will have a smaller free travel time as compared to the other photons. The \textit{mean free} time is a measure of the free travel time for a given photon, and we use each photon's mean free time (eq.~\ref{eq:mft_ithphoton}) and Monte Carlo techniques to identify and select the scattering photon. For any given photon, the probability of scattering is inversely proportional to its mean free time. Thus by inverting the mean free time $t_{\rm i}$ (see eq.~\ref{eq:mft_ithphoton}) we obtain a number $p_{\rm i}$ proportional to the likelihood for that photon to scatter. For $p_{\rm i}$ to be a probability distribution, we normalize the inverses obtained from all photons, by obtaining the norm $A$ --
\be
{\rm A} = \sum_{i = 1}^{\rm N_p} p_{\rm i} = \sum_{i = 1}^{\rm N_p} \frac{1}{t_{\rm i}}.
\ee
By normalizing the inverses using $\rm A$, we create a probability distribution amenable to Monte Carlo techniques. As each photon has a mean free time and an associated scattering probability, it is uniquely represented along the probability distribution by a certain range of values. We generate a uniformly distributed random variable in the interval [0,1) to identify which photon scatters.\\
The code follows a similar procedure to select the scattering lepton. To generate the probabilities for each lepton, we use the mean free time of interaction of the selected photon with the leptons in the plasma (given by eq.~\ref{eq:opacity_1pho_1lep_motion}). As a result, we obtain the inverses $p_{\rm i,j}$ and the new norm $\rm A_{\rm i}$-
\be
\rm A_{\rm i} = \sum_{j = 1}^{\rm N_e} p_{\rm i,j} = \sum_{j = 1}^{\rm N_e} \frac{1}{t_{\rm i,j}} 
\ee
As illustrated by the photon selection process, each normalized probability $\rm p_{\rm i,j}/{\rm A_{\rm i}}$ represents a range of values, thereby, identifying a unique lepton. We generate a uniform random variable in the interval [0,1), and thus identify which lepton among the $N_e$ leptons scatters with the selected photon.

\subsubsection{Particle Propagation}\label{sec:part_prop}
Between the scattering/collision events, the particles freely propagate for the duration of the time step. The position of any particle at time $t$ is given by-
\be
x^i(t) = x^i(t - t_{\rm coll}) + c\frac{p^i}{p^0}t_{\rm coll}
\ee
where $x^i(t - t_{\rm coll})$ denotes the $i^{th}$ position component of the particle at time $t - t_{\rm coll}$, $p^i$ and $p^0$ denote respectively the spatial and the zeroth component of the four--momenta of the particle.

\subsubsection{Periodic Boundary Condition}\label{sec:PBC}
As explained in \S~\ref{sec:part_prop}, once the time step is determined, the DynaMo code propagates and tracks the particles constituting the GRB fireball. Depending upon their momenta, particles can propagate to positions outside the simulated wedge resulting in the simulation losing particles. The radial thickness of the simulated fireball is determined by the innermost and outermost leptons, and so by the virtue of this definition, matter can never escape the simulated region radially. However, photons in the plasma can and are allowed to escape, as is discussed in  \S~\ref{sec:photonescape}. However, all particles can also propagate to regions outside the wedge through the lateral surfaces (as shown by particle A in Fig.~\ref{fig:diagram}). These laterally escaping particles are, in fact, still within the spherical fireball (albeit outside our current simulation) and will influence the evolution of the entire spherical fireball. Just as a particle escapes our simulated wedge to an adjacent region, a particle from an adjacent region can enter via these lateral surfaces. 
To correctly simulate a spherical fireball the DynaMo code implements a periodic boundary condition scheme to account for particles traveling across lateral boundaries. This scheme implicitly assumes that the total particle count of the fireball is constant. If any particle exits through the lateral surface of the simulated wedge making an angle $\varphi$ with the exiting surface, then a corresponding particle is inserted through the wedge surface opposite to the exit location, but at the same radius (see Fig.~\ref{fig:diagram}). The inserted (re-entering) particle makes the same angle $\varphi$ with the entering surface as the exiting particle made with the exit surface, and possesses the same energy. This scheme is similar to the periodic boundary condition for parallel surfaces (where the inclination angle between the surfaces is zero) where the particle momentum vectors at exit and entry surfaces are parallel and hence unrotated. As our simulated wedge has inclined surfaces, a lateral surface crossing requires the particle's three momenta be rotated by an angle of $2\theta_c$ (or twice the opening angle of the wedge). It should be noted that energy and total momentum magnitude of the particle remains constant during this scheme's implementation. However, the direction of three momentum does not remain constant.

\subsubsection{Compton Scattering}
The dynamical energy transfer mechanism in our simulations is Compton / Inverse Compton scattering. Once the particle selection step decides which pair of particles will be scattered, we use the Compton scattering algorithm (Klein--Nishina regime included) discussed in \cite{CL15} for the scattering event. Post scattering, we have a new four--momenta for both the scattered lepton and photon which we use to calculate a new time step for the next scattering.

\subsubsection{Baryon--Lepton Collisions}
In general, GRB fireballs are baryon loaded (\citealt{MesLagunaRees93}) making them significantly more massive than baryon--deficient fireballs. We wish to explore whether the baryons (protons) in these baryon--loaded fireballs can accelerate and attain relativistic speeds (just like electrons do). We ignore the photon--proton ($\gamma p$) scatterings because the photon energies have to be comparable to rest mass energy of protons ($\sim$ GeV) for the scattering process to transfer an appreciable amount of energy. To accelerate the protons the DynaMo code accounts for energy transfer between electrons and protons via electron--proton ($e^- p$) collisions. To simplify the physics and the collision process, the code does not carry out $e^- p$ collisions using the Coulomb cross section, instead it performs elastic collisions between electrons and protons. We note that our goal is to explore if protons can be accelerated to relativistic speeds, and not to simulate the $e^- p$ collision process to perfection. As our results demonstrate (see \S~\ref{sec:resanddiscuss}), the elastic collisions we use provide a fast and reliable method to mimic the energy transfer that would have occurred during a real collision using Coulomb cross section. Another simplification we make is employing pseudo protons to mimic the behavior of massive baryons. In order to accelerate protons (which are 1838 times the electron's mass) the code requires around 2000 additional photons for each proton in the plasma. To ensure that the simulations can be completed in a reasonable time frame, we employ pseudo protons of effective mass $m_{\rm eff} = 5 m_e$ instead of real protons that are 1838 times more massive. This is similar to the reduced ion-masses employed by multi-dimensional PIC simulations (see~\citealt{Spitkov2008}).\\
The Coulomb cross section for interaction between charged particles (like $e^- p$ interactions) is much larger than the cross section of $\gamma e^-$ interactions. Due to the attractive nature of the forces between these charged particles, there will be many $e^- p$ interactions for every $\gamma e^-$ interaction. In other words, for every Compton scattering event, there will be several elastic $e^- p$ collisions. 
As a consequence, between successive $\gamma e^-$ scatterings, the code ensures that each electron collides with a proton (for every $\gamma e^-$ scattering the number of $e^- p$ collisions equal the number of $e^- p$ pairs in the system).\\
To perform this elastic collision, the code begins by randomly selecting one $e^- p$ pair from the plasma. To simulate an $e^- p$ collision and the involved energy transfer, we first Lorentz transform to the proton's reference frame and compute the transformed four--momentum of the incoming electron. To calculate the transferred energies, we move to the center of momentum frame (COM frame) of the system. The advantage of employing this frame is that the collision here is always head on and the net three momenta of the $e^- p$ system, pre and post collision is always zero. We Lorentz transform the four--momenta of the colliding proton and electron to this frame, which completely specifies the pre--collision geometry. To specify the post--collision geometry, we randomly generate the polar angle $\theta'$ and the azimuthal angle $\phi'$ in the COM frame. Now, by employing conservation laws, we can obtain the four--momentum of the proton after collision.\\
The four--momenta thus obtained in the COM frame is de-boosted to get the momenta in the proton's frame and subsequently, in the lab frame. This provides us with the post--collision momenta of both the proton and the electron in the lab frame. As stated earlier, for every $\gamma e^-$ scattering event, the code collides all the $e^- p$ pairs once, leading to energy and momentum exchange among these particles. As a result, though the photons and the protons do not interact directly, the electrons act as an intermediary for transferring energy between radiation and baryons.

\subsubsection{Data Update and Photon Escape Condition}\label{sec:photonescape}
Once the code performs the scattering and collisions, it updates the positions and four--momenta of all particles involved in the simulation. After each time step the code computes the radius of the farthest and nearest lepton from the origin, thereby,  determining the fireball's outer and inner radii, respectively. To evaluate if a photon has escaped from the fireball, the code compares the radii of all photons with the fireball's outer radius. If any photon's radius exceeds the outer radius of the fireball, that photon is deemed to have escaped the fireball and is no longer a part of our simulation (it will not be involved in the time-step calculation nor the scattering process). The code runs until all the photons escape and are unable to further accelerate the fireball.

\section{Results and Discussion}\label{sec:resanddiscuss}
\begin{figure}
	\centering
	\includegraphics[width=1.0\linewidth]{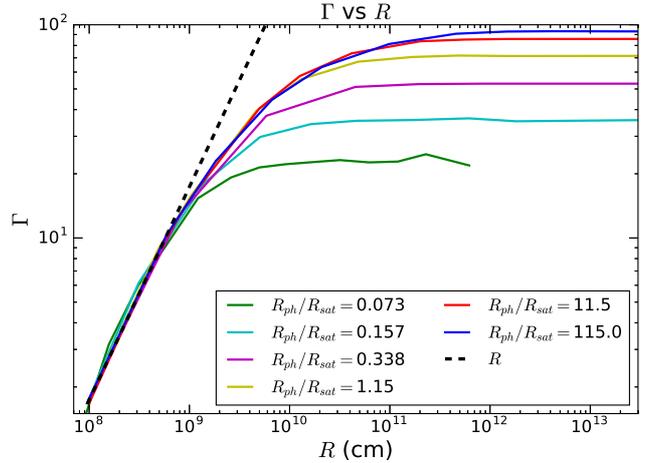}
	\caption{Evolution of Lorentz factor with radius for fireballs characterized by different initial opacities. 
	The legend displays the $R_{\rm ph}/R_{\rm sat}$ value for each fireball and the corresponding color representing it.}
	\label{fig:grad-v-rrad-diff-sims}
\end{figure}
\begin{figure}
	\centering
	\includegraphics[width=1.0\linewidth]{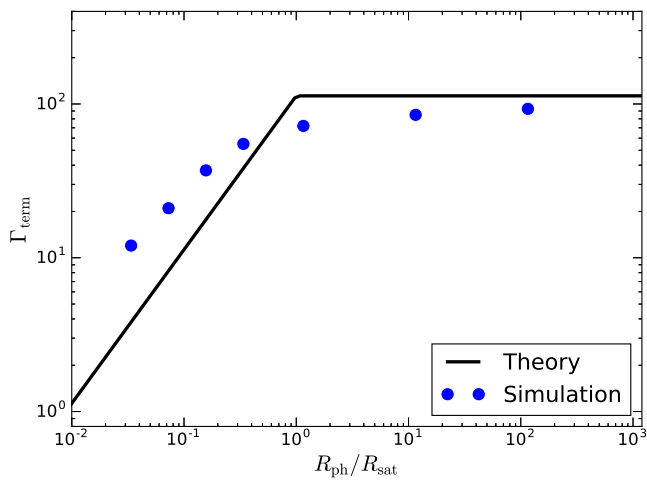}
	\caption{Theoretical and simulated values of terminal Lorentz factor $\Gamma_{\rm term}$ plotted against different $R_{\rm ph}/R_{\rm sat}$ ratios. The black curve plots the theoretical value of the terminal bulk Lorentz factor (obtained from the fireball model). The blue points represent the results obtained from our DynaMo simulations.}
	\label{fig:gsat-v-rphrsat}
\end{figure}
We have performed simulations of Compton scattering induced -- radiative acceleration and expansion of a fireball, with and without baryon loading. Figure~\ref{fig:grad-v-rrad-diff-sims} shows the radial evolution of the bulk Lorentz factor of baryon--deficient fireballs for several ratios of $R_{\rm ph}/R_{\rm sat}$ (which serves as a proxy for the opacity). A detailed calculation of $R_{\rm ph}/R_{\rm sat}$ can be found in Appendix~\ref{sec:calcRphRsat}. These results confirm the idea that greater the opacity of a plasma, the greater the terminal Lorentz factor achieved by the plasma due to acceleration by embedded radiation. Alternatively, as opacity indicates the number of scatterings occurring in a medium, these results confirm that an increasing number of scatterings is necessary for a plasma to continue accelerating radiatively by converting its internal energy into bulk motion. The evolution of our simulated fireballs is in strong agreement with the fireball model's proportionality relation $\Gamma \propto R$, during the radiation--dominated acceleration phase. In addition to $\Gamma \propto R$, the simulated fireballs also follow the relation $T^{\prime} \propto R^{-1}$. The radial evolution of comoving temperature and other code tests can be found in Appendix~\ref{sec:sub_codetest}.
\subsection{Transition Regime and Lorentz factor turnover}
An interesting and new feature that emerges from Fig.~\ref{fig:grad-v-rrad-diff-sims} is the transition of the simulated Lorentz factors from the $\Gamma \propto R$ (radiation dominated) to $\Gamma \propto R^0$ (matter dominated) regime. As we pointed out in \S~\ref{sec:sub_fireball_theory}, departure from $\Gamma \propto R$ regime occurs because fireball's energy density is no longer dominated by radiation. Further, we note that the magnitude of curvature or turnover of the Lorentz factor (the measure of how rapidly $\Gamma$ transitions from $\Gamma \propto R$ to $\Gamma \propto R^0 $) is opacity dependent. This implies that more opaque fireballs (e.g, blue curve in Fig.~\ref{fig:grad-v-rrad-diff-sims}) transition more smoothly and gradually (with larger curvature) as compared to less opaque fireballs (e.g., green curve). A consequence of this gradual transition is an increase in the radial extent of the fireballs' transition zone, which spans several orders of magnitude (it is also larger in extent than the $\Gamma \propto R$ regime).
   
\subsection{Post--Photospheric Acceleration Phase}
Figure~\ref{fig:gsat-v-rphrsat} plots the terminal Lorentz factor of several fireballs against their corresponding $R_{\rm ph}/R_{\rm sat}$ values. For a given $R_{\rm ph}/R_{\rm sat}$ ratio, the black curve depicts the theoretically calculated terminal Lorentz factor ($\Gamma_{\rm Th}$). The blue points show the terminal Lorentz factor obtained by DynaMo simulations (let us call them $\Gamma_{\rm Sim}$). For $R_{\rm ph}/R_{\rm sat}<1$, the fireball model predicts that all radiation escapes at the photoshere and the Lorentz factor saturates, i.e., $\Gamma_{\rm Th} = \Gamma_{\rm ph}$ (see \S~\ref{sec:sub_fireball_theory} and Appendix~\ref{sec:calcRphRsat}). However, Fig.~\ref{fig:gsat-v-rphrsat} shows that for $R_{\rm ph}/R_{\rm sat}<1$ simulated Lorentz factors exceed the corresponding theoretical values, i.e., $ \Gamma_{\rm Sim} > \Gamma_{\rm Th}$. Physically, this excess represents an acceleration phase occurring \textit{after} the theoretical photosphere. \cite{Paczynski1990} was able to  identify this excess acceleration regime but was unable to analytically study this regime using the obtained solutions. This is due to the fact that to obtain the outflow solutions the flow velocity beyond the photosphere was assumed to be constant. Our results show (and this is also mentioned in \citealt{Paczynski1990}) that this assumption breaks down for outflows that remain radiation dominated until their photosphere (thereby experiencing acceleration beyond their photospheres).\\\\
The reason underlying this excess acceleration lies in the definition of opacity. The opacity of a medium is a probabilistic quantity and can only provide information regarding the probability of escape of a photon (\citealt{Peer2k8Tempevol}). However, the theoretical values (such as the $\Gamma_{\rm Th}$ at the photospheric radius) are calculated assuming that all the photons in the fireball escape at $\tau = 1$ (\citealt{ReesMesz2005dissphotos}) and do not account for the opacity's probabilistic nature. 
By using the particle tracking feature of the DynaMo code we find that even for optically thin fireballs (i.e., $\tau \leq 1$) a fraction of photons are still trapped, which concurs with the probabilistic nature of opacity. These trapped photons (that are unaccounted for in the fireball model) continue scattering beyond the photosphere and are the reason why $\Gamma_{\rm Sim}$ exceeds $\Gamma_{\rm Th}$. It is interesting to note that the smaller the value of $R_{\rm ph}/R_{\rm sat}$ the larger the discrepancy between simulated and theoretical values. This can be attributed to the fact that the (trapped) radiation energy density is larger for smaller $R_{\rm ph}/R_{\rm sat}$ fireballs, which leads to greater acceleration beyond the photosphere. As $R_{\rm ph}$ grows, the radiation energy density decreases and it becomes harder for less energetic, trapped photons to provide that extra push, causing $\Gamma_{\rm Sim}$ to converge to $\Gamma_{\rm Th}$. On the fireball phase diagram (Fig.~\ref{fig:fbphasespace}), the post--photospheric acceleration phase occurs during the transition from quadrant 1 to 3 via 2.
\subsection{Thompson--Dominated Acceleration Phase}
According to the standard fireball model, for $R_{\rm ph}/R_{\rm sat} \geq 1$ the fireball enters the saturation phase and thereafter coasts at $\eta$ (see \S~\ref{sec:sub_fireball_theory} and Appendix~\ref{sec:calcRphRsat}). For $R_{\rm ph}/R_{\rm sat} \geq 1$ and in contrast to the results of the post--photospheric acceleration phase, Fig.~\ref{fig:gsat-v-rphrsat} shows that $\Gamma_{\rm Sim} \leq \Gamma_{\rm Th}$. Another interesting observation is that the for increasing $R_{\rm ph}/R_{\rm sat}$ values the discrepancy between simulated and theoretical values is reduced, and $ \Gamma_{\rm Sim} \to \Gamma_{\rm Th} = \eta $ asymptotically. These features can be attributed to the fact that 1) at this stage these fireballs are matter dominated and 2) Comptonization is inefficient for transferring energy from low energy photons to leptons. The matter--dominated fireball phase implies that the average lepton energy is significantly larger than the average photon energy. As seen from the comoving frame, the photon energies are significantly lower than the rest mass energy of electrons and scatterings here occur in the \textit{Thomson regime}. In order to extract the final remnants of the significantly smaller photon energies and reach the maximum allowed Lorentz factor, a large number of (Compton) scatterings are required (\citealt{CL15}). Our results show that once fireballs become matter dominated, Comptonization is not efficient in completely converting internal energy into bulk motion. On the phase--space diagram shown in Fig.~\ref{fig:fbphasespace}, the fireball encounters this phase as it evolves from quadrant 1 to 3, transitioning via quadrant 4. It is this Thomson--dominated phase that is responsible for the gradual fireball acceleration, the gradual $\Gamma$ turnover (which becomes increasingly smooth for higher opacities) and the large extent of the transition phase. Thus, optically thick fireballs that become matter dominated accelerate gradually, and require extremely large opacities (or an extremely large number of scatterings) to approach $\eta$.
\subsection{Baryon Loading}
\begin{figure}
	\centering
	\includegraphics[width=1.0\linewidth]{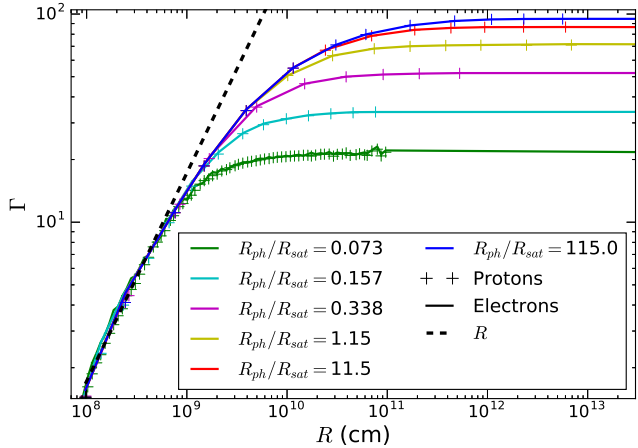}
	\caption{Same as Fig.~\ref{fig:grad-v-rrad-diff-sims} but with fireballs that are baryon loaded. Also plotted is the Lorentz factor evolution of protons (represented by $+$ markers).}
	\label{fig:gradvsrrad_epro}
\end{figure}
Figure~\ref{fig:gradvsrrad_epro} plots the radially evolving Lorentz factors of baryon--loaded fireballs, with each colored curve corresponding to a unique value of $R_{\rm ph}/R_{\rm sat}$. The fireball evolution is again in strong agreement with the fireball model's proportionality relations (see eq~\ref{eq:fireball_acc}) during the radiation--dominated acceleration phase. We observe again that fireballs with larger $R_{\rm ph}/R_{\rm sat}$ attain larger Lorentz factors. Thus, both Figs.~\ref{fig:grad-v-rrad-diff-sims} \&~\ref{fig:gradvsrrad_epro} confirm the idea that larger opacity leads to larger acceleration, independently of the fireball's baryonic content.\\
The inclusion of baryons increases the effective mass of the plasma and in comparison to baryon--free plasmas requires more photons for acceleration. This leads to an increase in computational time and memory consumption. Since both baryon--devoid and baryon--loaded plasmas have extremely similar Lorentz factor evolution and terminal values at saturation, we think it is better to simulate just baryon--free fireballs as they require less computational time and resources.\\

\subsection{Expression from curve fitting}\label{sec:sub_curvefitting}
\begin{figure}
	\centering
	\includegraphics[width=1.0\linewidth]{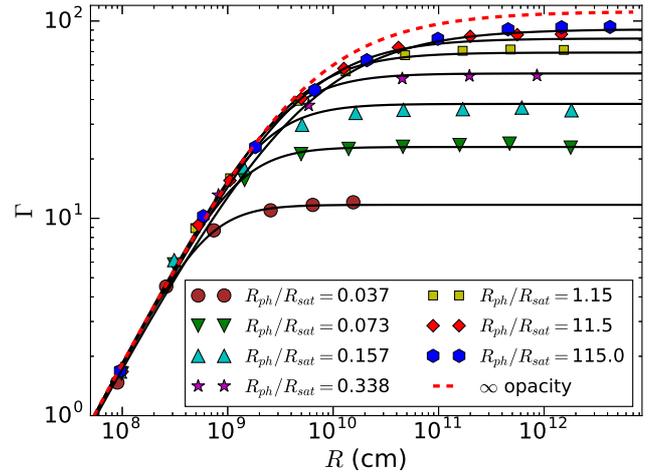}
	\caption{Lorentz factor evolution with increasing radii of fireballs characterized by different initial opacities. The differently shaped and colored markers plot eq.~\ref{eq:analyticalexp}, which is obtained by curve fitting (see  \S~\ref{sec:sub_curvefitting}). The solid lines plot the simulation results. The red dashed line is an analytical result (see eq.~\ref{eq:altmodelGamma}) obtained from an infinite opacity fireball (discussed in \S~\ref{sec:sub_fireball_theory}).}
	\label{fig:analyticmod}.
\end{figure}
In this section, we present and discuss the analytical expressions obtained by fitting the simulated data depicted in Fig.~\ref{fig:grad-v-rrad-diff-sims}. Each simulation in Fig.~\ref{fig:grad-v-rrad-diff-sims} is characterized by unique initial outflow conditions -- value of injection radius $R_0$, maximum possible bulk Lorentz factor $\eta$ (attained when all internal energy is converted to bulk motion) and the unique parameter $\sigma$ (a measure of opacity of the plasma; its relation to $R_{\rm ph}/R_{\rm sat}$ is described in detail in Appendix~\ref{sec:calcRphRsat}). The expression we use to model the radial evolution of the Lorentz factor $\Gamma(R)$ is (the expression used has a form similar to \citealt{Beuer1999_analyfit}) -- 
\be
\Gamma(R) = \frac{\Gamma_{\infty}(\sigma, \eta)}{\left[1 + \left( \frac{r_{\rm acc}(\sigma, \eta)}{r} \right)^{s(\sigma)} \right]^{\frac{1}{s(\sigma)}}},\label{eq:analyticalexp}
\ee
which depends primarily on three parameters, which in turn are functions of the initial outflow conditions. These three parameters are --
\bite
\item $s(\sigma) = 2.53 - 0.1796 \log_{10} \sigma  $
\item $\Gamma_{\infty} (\sigma, \eta) = 10^{\left(\log_{10} \eta \left[1 - \exp(-0.43 log_{10} \sigma) \right]^{3.3}\right)}$
\item $r_{\rm acc} (\sigma, \eta, R_0) = 0.54 R_0 \eta$
\eite
Fig.~\ref{fig:analyticmod} compares the Lorentz factors calculated from the numerical expression obtained via curve fitting (i.e., eq.~\ref{eq:analyticalexp} and depicted by markers) and the simulation results (represented by the solid lines). As the markers and the solid lines show, the numerical and simulated results are in good agreement with each other. The dashed red line is the Lorentz factor calculated from eq.~\ref{eq:altmodelGamma}, which was obtained from the hydrodynamic evolution of an infinite opacity fireball (see \S~\ref{sec:sub_fireball_theory}). The evolution of the red dashed line is very similar to our simulated, optically thickest fireball during the radiation--dominated and transition phases. 
\section{Summary and Conclusions}\label{sec:conclu}
\begin{figure}
	\begin{center}
		\includegraphics[width=1.0\linewidth]{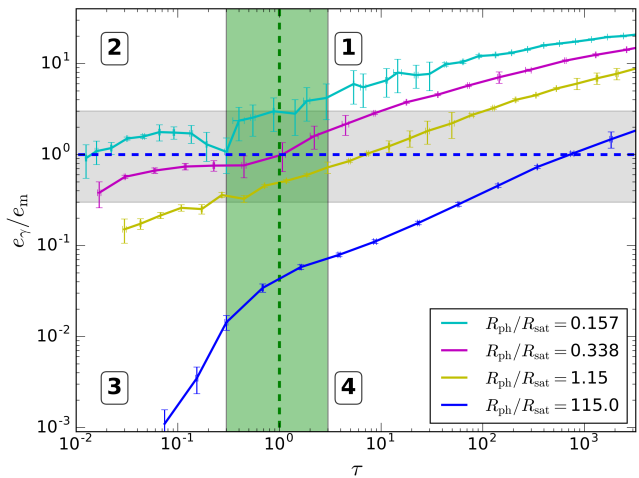}
		\caption{An updated version of Fig.~\ref{fig:fbphasespace} with the DynaMo code's simulation results in context. The differently colored curves represent fireballs starting with different initial opacities. Fireballs that start comparatively optically thin ($R_{\rm ph}/R_{\rm sat} < 1$) evolve from quadrant 1 to 3 via 2. Comparatively optically thick fireballs ($R_{\rm ph}/R_{\rm sat} \geq 1$) evolve from quadrant 1 to 3 via quadrant 4. All curves eventually drop because only photons trapped inside the fireball contribute to the radiation energy density $e_{\gamma}$. The Lorentz factor evolution across the entire phase space is given by eq.~\ref{eq:analyticalexp}.}\label{fig:Dyanmofbphasespace}
	\end{center}
\end{figure}
The mechanism that launches and accelerates the ultra--relativistic GRB jets is a mystery. The outflows in GRBs can be magnetically accelerated if the jets are strongly magnetized, i.e., the fields carry a significant amount of total energy, in the form of Poynting flux which is converted into (bulk) kinetic energy (\citealt{Lyuba2010_PoynftoKE};\citealt{KumaZhang15}). Several works have shown that these jets can be accelerated adiabatically via magnetic pressure, e.g. (\citealt{Conto1995_PlasmaGun}; \citealt{Granot2k11_ImpulAccMagnetOutfl}). Under certain magnetic field configurations, magnetic reconnection and dissipation can also accelerate GRB outflows (\citealt{DrenkSprui02}; \citealt{Drenk2002_GRNaccPoyntDissp}; \citealt{ZhangYan2011_ICMART}). For magnetically accelerated outflows, the proportionality relation between the Lorentz factor and the outflow radius grows as $\Gamma \propto R^{1/3}$ (\citealt{KumaZhang15}), which is distinctly different from the Lorentz factor evolution of radiation dominated outflows (eq. \ref{eq:fireball_acc}). As several authors have investigated magnetized outflows, in our work we focused on an alternate launching and acceleration mechanism for outflows, i.e., radiation.\\
In this paper, we have studied the bulk kinematic evolution of a radiation--driven GRB fireball via Monte Carlo simulations. We have presented our novel Dynamic Monte Carlo code (DynaMo) code, which we use to self--consistently simulate, the Compton scattering--induced expansion and acceleration of a GRB fireball. 
Earlier works have studied particular phases of fireball evolution analytically (\citealt{MesLagunaRees93}; \citealt{Piran1999}). The analytical approximations do not properly address the question of self consistent fireball evolution, especially when 1) the radiation energy density becomes comparable to or less than the fireball's rest mass energy density and/or 2) the fireball becomes optically thin (the case when $R_{\rm ph} < R_{\rm sat}$). 
Fig.~\ref{fig:Dyanmofbphasespace} is an updated version of the GRB phase space diagram (Fig.~\ref{fig:fbphasespace}) with the evolution of DynaMo code's simulated fireballs plotted onto the diagram. We have studied fireball evolution across all regimes, including the transition regimes that are the highlighted regions in phase diagrams (see Figs.~\ref{fig:fbphasespace} \&~\ref{fig:Dyanmofbphasespace}). We summarize our results as follows --
\bite
\item The evolution of a GRB fireball (or any outflow) can be summarized by Fig.~\ref{fig:Dyanmofbphasespace}. All radiation--dominated and optically thick GRB fireballs start in the upper--right quadrant (quadrant 1) and evolve towards quadrant 3 (optically thin, matter dominated). Depending on where the fireball originates within quadrant 1, it will enter quadrant 3 via paths through quadrants 2 or 4.
\item We have investigated the effect of initial opacity on fireball evolution by simulating several fireballs, each starting with a different initial opacity (as differentiated by the colored curves in Figs.~\ref{fig:grad-v-rrad-diff-sims} and~\ref{fig:gradvsrrad_epro}). Our results are in agreement with the fireball model, as optically thick fireballs achieve higher Lorentz factors, and saturate at larger radii than comparatively thin ones. 
\item We have also investigated the effects of baryon loading of fireballs on the Lorentz factor evolution, as shown by Fig.~\ref{fig:gradvsrrad_epro}. Our results show identical evolutionary behavior of baryon--deficient and baryon--loaded fireballs (Figs.~\ref{fig:grad-v-rrad-diff-sims} and~\ref{fig:gradvsrrad_epro} evolve similarly). Being baryon loaded, these fireballs require more photons to accelerate the additional mass and hence require more computational time and resources. Our results strongly suggest that the evolution of baryon--loaded fireballs can be accurately predicted from the significantly faster and less memory intensive, baryon--deficient fireball simulations. This can also be realized with the aid of eq.~\ref{eq:etadef}, as the mass term in the denominator does not differentiate between baryonic or leptonic mass.
\item A remarkable result that can be seen from both Figs.~\ref{fig:grad-v-rrad-diff-sims} and~\ref{fig:gradvsrrad_epro} is the existence of a transition regime, with an opacity dependent radial extent. Occuring between the radiation--dominated acceleration phase (where $\Gamma \propto R$) and the phase where the Lorentz factor flattens ($\Gamma \propto R^0$), the transition phase has 1) a significant radial extent (which exceeds the radial extent of $\Gamma \propto R$ phase) and 2) a curvature / turnover radius which gradually increases with increasing opacity. In Fig.~\ref{fig:Dyanmofbphasespace}, this regime begins within the highlighted regions, where opacity approaches unity and/or radiation energy no longer dominates the fireball energy density.\\
\item The simulation results show the existence of a post--photospheric acceleration phase (see Fig.~\ref{fig:gsat-v-rphrsat}) as predicted by \cite{Paczynski1990}, during the previously unexplored $R_{\rm ph} < R_{\rm sat}$ regime. This phase is encountered by the fireball as it travels from quadrant 1 to 3 via 2 (see Fig.~\ref{fig:Dyanmofbphasespace}). In this regime, simulated fireballs' terminal Lorentz factors are found to be larger than the theoretical Lorentz factors obtained using the fireball model. In other words, the simulated fireballs continue accelerating beyond the theoretical photosphere. We attribute this post--photospheric acceleration phase to the energetic photons still trapped in the plasma. The particle tracking feature informs us that 1) not all photons escape at the theoretical photosphere (when $\tau \sim 1$) and 2) photons at the these small radii are sufficiently energetic. These trapped, energetic photons continue to scatter and accelerate the fireball beyond the model's theoretical values.\\
\item Another interesting result is the existence of the Thomson--dominated acceleration phase, detected in the regime $R_{\rm ph} \geq R_{\rm sat}$ (see Fig.~\ref{fig:gsat-v-rphrsat}). Conversely to the post--photospheric acceleration phase, this phase is characterized by the theoretical Lorentz factor exceeding the simulated terminal Lorentz factor. On the phase space diagram (Fig.~\ref{fig:Dyanmofbphasespace}), this phase arises as the plasma evolves from quadrant 1 to 3 via 4. This phase can be attributed to the fact that in the optically thick and matter--dominated regime, the average energy of matter is greater than the average photon energy. As a result, energy transfer per scattering is low (Thomson scattering) and for Comptonization to transfer significant energy from radiation to matter, a large number of scatterings is required. The Thomson--dominated scattering phase is responsible for the gradual acceleration and the large transition / gradual turnover phase for fireballs with large $R_{\rm ph} \geq R_{\rm sat}$.
\eite
We have shown that radiation can accelerate fireballs beyond the photosphere (post--photospheric acceleration -- see \S~\ref{sec:resanddiscuss}) and to Lorentz factors larger than previously estimated. A powerful implication of our DynaMo code is that the photosphere cannot be assigned a single value or a radius (and not all radiation escapes at this value), instead it corresponds to a region or volume from where photons gradually escape (\citealt{Peer2k8Tempevol}; \citealt{Belob2011}; \citealt{Mcrat16}). While escaping through this volume, the radiation scatters and accelerates the plasma leading to larger than expected acceleration. Another consequence of our results is the re--definition of saturation radius. The saturation radius defined using eq.~\ref{eq:rsat} does not hold true if the energy density is not dominated by radiation. Our results show that at the end of the radiation--dominated acceleration phase, an opaque fireball can still accelerate but only gradually (Thomson--dominated acceleration phase). As a result, the fireball can attain $\eta$ only for extremely large opacities (or equivalently a large number of scatterings) and at radii significantly larger than the saturation radius (see Fig.~\ref{fig:gsat-v-rphrsat}).\\
We quantify the radial evolution of the Lorentz factor by fitting the simulated data using $\chi^2$ technique. The analytical expression obtained (see eq.~\ref{eq:analyticalexp}) is parametrized using the fireball's initial opacity. The advantage of this expression is that it can successfully capture the evolution of the GRB fireball across its entire phase space. This includes the transition regimes as well as the fireball evolution in quadrant 2. As an example, by supplying the relevant input parameters (such as the initial opacity) to eq.~\ref{eq:analyticalexp}, the Lorentz factor at any radius can be obtained for an evolving GRB fireball.\\
Our simulated fireballs are hot and dense enough that electron--positron pair processes can become important in changing the photon and lepton counts. At the beginning of the fireball evolution the changing particle count only serves to increase the opacity which decreases the escape probability of the photons. As a result, radiation and matter remain in equilibrium as the fireball expands, cools and eventually reaches a temperature where pair processes become unimportant and thus can be ignored. In all our simulations, pairs become irrelevant long before radiation escapes the plasma and thus pair processes are not accounted for.\\
The results obtained in this paper will be useful for studying photospheric emission from GRBs. Though this paper only focuses on GRB fireballs, the DynaMo code can be used for studying radiative acceleration in relativistic outflows from other astrophysical environments, such as AGNs and microquasars. The DynaMo code's particle tracking feature allows the user to not only obtain the spectrum of the escaping radiation, but also determine when and where the photons escape the outflow. The code can thus produce time--resolved spectra and light curves from GRB fireballs, which will be the subject of future publications.\\

\section*{Acknowledgements}
We thank the anonymous reviewer for suggestions to improve the manuscript. This work made use of NumPy (\citealt{Numpy}) for computation, and Matplotlib (\citealt{Matplotlib}) for preparing figures. This work was supported in part by NASA Swift Grant NNX15AG96G.




\bibliographystyle{mnras}
\bibliography{firstbib} 




\appendix
\section{Time Step Calculation Details}\label{sec:Appdx_timestep}
In this section we detail our calculation of the DynaMo code's dynamic time step. We begin with a discussion of the opacity experienced by a photon immersed in a collection of scatterers at rest. We use this opacity to calculate the mean free path for a single photon. We then extend this analysis to calculate the opacity of a medium containing several photons and the mean free time of this population of photons.\\
We then detail how to incorporate the motion of scatterers into the opacity and mean free time calculations. As before, we derive the expression for mean free time for a single photon immersed among scatterers in motion. We use this result to find the mean free time for a population of photons in a moving medium.
\subsection{For scatterers at rest}\label{sec:tstep_appdx_scatrest}
Using eq.~\ref{eq:opacity_basic} (which is valid for scatterers at rest) as the defining equation of opacity, the mean free path ($l_{\rm mf}$) traversed by the photon can be computed to be --
\be
l_{mf} = \frac{1}{n \sigma}\label{eq:mfp_basic}.
\ee
We take advantage of the constancy of the speed of light to obtain the mean free time $t_{\rm mf}$ as -- 
\be
t_{\rm mf} = \frac{1}{n c\sigma}\label{eq:mft_basic}.
\ee
The reader should note that this is the \textit{mean} or \textit{average} time that the photon travels between scatterings. For a medium containing $N_{\rm p}$ photons, the total infinitesimal opacity can be obtained by summing the individual contribution from each photon --
\be
d\tau_{\rm Pop} = \sum_{i=1}^{N_{\rm p}} d\tau_i = \sum_{i=1}^{N_{\rm p}} n \sigma dl_i = n \sigma \sum_{i=1}^{N_{\rm p}} dl_i = n \sigma l_{\rm mf, Pop} N_{\rm p}, \label{eq:opacity_Npho_app}
\ee
where the subscript $\rm Pop$ is used for population and $l_{\rm mf, Pop}$ denotes the mean free path for the photon population. Similar to eq.~\ref{eq:mfp_basic}, we can express the mean free path of the entire population as -- 
\be
l_{\rm mf, Pop} = \frac{1}{n \sigma N_{\rm p}} = \frac{l}{N_{\rm p}} \label{eq:mfp_basic_Npho}.
\ee
The mean free time for this population can be computed to be --
\be
t_{\rm Pop} = \frac{l_{\rm mf, Pop}}{c} = \frac{1}{N_{\rm p} c n \sigma } = \frac{t_{\rm mf}}{N_{\rm p}}.\label{eq:mft_1pho_manylep}
\ee
\subsection{For scatterers in motion}\label{sec:tstep_appdx_scatmot}
In this section we derive the opacity and mean quantities when the scatterers in the medium are moving (i.e., the medium itself is moving) and compare them with the same quantities obtained when the scatterers are at rest. We begin with the case of just a single photon and then extend our calculation to a system containing multiple photons.\\
Consider the case where each of the scatterers travel at a unique velocity and hence contribute uniquely to the opacity. The differential opacity seen by a photon due to scatterers traveling 
at a distinct velocity given by $\vec{\beta_{\rm k}}$ is (see \citealt{AbramNovikPaczy91}) --
\bea
d\tau_{\rm k} &=& n_{\rm k} \sigma (1 - \vec{\beta_{\rm k}}.\hat{r}) dl \\
d\tau_{\rm k} &=& n_{\rm k} \sigma (1 - \beta_{\rm k} \cos \theta_{\rm k}) dl
\eea
where $n_{\rm k}$ denotes the number density of scatterers traveling in the $k^{\rm th}$ direction with a speed $\beta_{\rm k}$, $\hat{r}$ is the unit vector along the direction of propagation of the photon and $\theta_{\rm k}$ is the angle between the photon's direction of propagation and $k^{\rm th}$ direction.\\
The total opacity as observed by the photon will be the sum of the opacities due to each individual scatterer and can be written as --
\bea
d\tau = \sum_{k} \tau_{\rm k} &=& \sum_{k} n_{\rm k} \sigma (1 - \vec{\beta_{\rm k}}.\hat{r}) dl \\
d\tau = \sum_{k} \tau_{\rm k} &=& \sum_{k} n_{\rm k} \sigma (1 - \beta_{\rm k} \cos \theta_{\rm k}) dl \label{eq:opacity_1pho_manylep_moving}
\eea
where $\sum_{k}$ denotes the sum over the directions and speeds associated with the scatterers. These expressions can be used to write the mean free time experienced by a photon traveling through a moving medium as --
\be
t_{mf} = \frac{1}{\sum_{k} n_{\rm k} c \sigma (1 - \beta_{\rm k} \cos \theta_{\rm k})}.\label{eq:mft_motion_1pho}
\ee
The reader should note that on comparison with a medium at rest, the photon observes a decrease in opacity (and an increase in the mean free time) if it moves along the direction of motion of the scatterer. For motion anti-parallel to the scatterer's motion, an increase in opacity is experienced and as a consequence, the mean free time is reduced.\\ 
If the total number of scatterers (say leptons) is fixed and denoted by the number $N_e$, we can re-write eq.~\ref{eq:opacity_1pho_manylep_moving} by eliminating $n_k$ as --
\be
d\tau = \frac{\sum^{N_e}_{k} \sigma (1 - \beta_{\rm k} \cos \theta_{\rm k})}{V} dl,  
\ee
where $V$ denotes the volume occupied by the scatterers. Re-writing the mean free time obtained in eq.~\ref{eq:mft_motion_1pho}, we obtain --
\be
t_{\rm mf} = \frac{l_{\rm mf}}{c} = \frac{V}{\sigma c \sum_{k=1}^{N_e} (1 - \beta_{\rm k} \cos \theta_{\rm k})}. \label{eq:mft_1pho_manylep_motion}
\ee
Let us now extend the above analysis to the most general scenario -- a plasma containing $N_p$ photons and $N_e$ leptons (which play the role of scatterers), and each lepton can be in motion. Similar to the calculation performed in Appendix~\ref{sec:tstep_appdx_scatrest}, the opacity experienced by the $i^{\rm th}$ photon due to \textit{any} lepton traveling along the $j^{\rm th}$ direction is --
\be
d\tau_{\rm i,j} = \frac{\sigma (1 - \beta_{\rm j} \cos \theta_{\rm ij})}{V} dl. \label{eq:opacity_1pho_1lep_motion}
\ee
Using eqs.~\ref{eq:mft_1pho_manylep_motion} and~\ref{eq:opacity_1pho_1lep_motion}, the mean free time for the interaction between $i^{\rm th}$ photon -- $j^{\rm th}$ lepton pair becomes --
\be
t_{\rm i,j} = \frac{l_{\rm ij}}{c} = \frac{V}{\sigma c(1 - \beta_{\rm j} \cos \theta_{\rm ij})}. \label{eq:mft_1pho_1lep_motion}
\ee
Using eq.~\ref{eq:mft_1pho_manylep_motion} and~\ref{eq:mft_1pho_1lep_motion}, the mean free time experienced by the $i^{\rm th}$ photon as it interacts with all $N_e$ leptons is --
\be
t_{\rm i} = \frac{V}{\sum_{j=1}^{N_e} \sigma (1 - \beta_{\rm j} \cos \theta_{\rm ij}) c} = \frac{V}{\sigma c}\frac{1}{\sum_{j} (1 - \beta_{\rm j} \cos \theta_{\rm ij})} = \frac{1}{\sum_{j} \left(\frac{1}{t_{\rm ij}}\right)}.\label{eq:mft_ithphoton}
\ee
Most generally, the net infinitesimal opacity of the photon population can be written as --
\be
d\tau_{\rm net} = \sum_{i=1}^{N_p} d\tau_{\rm i} =\sum_{i=1}^{N_p}  \sum_{j=1}^{N_e} \frac{\sigma (1 - \beta_{\rm j} \cos \theta_{\rm ij}) dl}{V}. \label{eq:opacity_net_pop_motion}
\ee
The use of eqs.~\ref{eq:mft_1pho_manylep_motion} and~\ref{eq:opacity_net_pop_motion} provides us with the mean free time for the entire population, which is --
\be
t_{\rm mf, Pop} = \frac{V}{\sigma c}\frac{1}{\sum_i \sum_{j} (1 - \beta_{\rm j} \cos \theta_{\rm ij})} = \frac{1}{\sum_i \sum_{j} \left(\frac{1}{t_{\rm ij}}\right)} = \frac{1}{\sum_i \left(\frac{1}{t_{\rm i}}\right)}.\label{eq:mft_pop_motion}
\ee
\section{Radius and Bulk Lorentz Factor Calculation}\label{sec:COMcalc}
As stated in \S~\ref{sec:sub_fireball_theory}, in order to characterize fireball evolution the bulk motion parameters are required. These parameters include the fireball's radius $R$, bulk Lorentz factor $\Gamma$ and the temperature $T^{\prime}$ measured in the frame comoving with the fireball. In this section we illustrate how we compute the fireball's radius and bulk Lorentz factor from the individual positions and momenta of the matter particles (leptons). In other words, we show how to find the radius and velocity of the Center of Momentum (COM) frame associated with the system of particles.\\
As the simulated fireball is a thin--shell, its constituent particles are distributed within the shell's finite width. Let us denote the $i^{th}$ particle's position as $r_i$ and velocity $v_i$. Generally, the particles do not travel along the radial direction (which lies along the $\hat{z}$ direction as explained in \S~\ref{sec:theory&meth}) because our simulation's origin is not coincident with each particle's position. As a consequence, we can resolve each particle's net velocity parallel and perpendicular to the radial vector at the particle's location. The radial components will contribute to the bulk motion whereas the non--radial components will provide us with a means to measure the temperature of the particles. The radial component of the velocity can be written as --
\be
v_{\rm rad,i} = \vec{v_i}.\hat{z}\label{eq:velocityradial_1particle}.
\ee
In general, the radial direction at each particle's position $\vec{r_i}$ is  different. However, if the angular size of the shell (in other words, the opening angle of our simulated fireball) is small then the positions of all particles lie within a narrow cone, and their radial vectors are approximately aligned. Using eq.~\ref{eq:velocityradial_1particle}, we can compute the radial Lorentz factor as --
\be
\gamma_{\rm rad,i} = \frac{1}{\sqrt{1 - \frac{v^2_{\rm rad,i}}{c^2}}}.
\ee
The radius of the fireball $R$ (or the radius of the COM) can be computed from the positions of the constituent particles as --
\be
R = \frac{\sum_i \gamma_{\rm rad,i} \vec{r_i}.\hat{r_i} }{\sum_i \gamma_{\rm rad,i}}.
\ee
To calculate the bulk Lorentz factor $\Gamma$, we start with the equation for total momentum of the system which gives us --
\be
\Gamma M V = \sum_i \gamma_{\rm rad,i} m_i v_{\rm rad,i}
\ee
where $V = c \sqrt{1 - 1/\Gamma^2}$. This equation can be simplified to --
\be
\Gamma \beta = \frac{\sum_i \gamma_{\rm rad,i} m_i \beta_{\rm rad,i}}{M} = \frac{\sum_i \gamma_{\rm rad,i} \beta_{\rm rad,i}}{N}, 
\ee
where we $\beta = V/c$, $\beta_{\rm rad,i} = v_{\rm rad,i}/c$ and $M = \sum^N_i m_i = N m$. Using the relation $\left(\Gamma \beta \right)^2 = \Gamma^2 - 1$, we can find the bulk Lorentz factor from the radial Lorentz factors of individual particles as --
\be
\Gamma = \sqrt{\left(\frac{\sum_i \gamma_{\rm rad,i} \beta_{\rm rad,i}}{N} \right)^2 + 1}.
\ee

\section{Comoving Temperature Calculation}
As the fireball accelerates and expands, it uses internal energy to fuel its bulk kinetic motion. The temperature of an object is a measure of the internal energy (attributed to random kinetic energy) content of that object. The net motion of the particles forming the fireball, in particular, the matter (e.g., leptons) -- is a composite of bulk motion (due to outward directed radiative acceleration) and random motion (due to temperature). Thus, determination of particles' temperature depends upon the particles' random speeds which can be correctly calculated only after accounting for their bulk velocity. As a result, we need to calculate the velocities of the particles as observed in the comoving frame and compute the temperature in that frame.\\
As discussed in section~\ref{sec:jet_init}, the fireball is a thin spherical shell that is expanding radially outward. The constituent particles of this shell are distributed randomly within the shell and interact with other particles in the shell. 
The radial speed of the center of mass can be computed as --
\be
V_{\rm COM} = \frac{\sum_i \gamma_i m_i v_{i, rad}}{\sum_i \gamma_i m_i}, 
\ee
where $v_{\rm i, rad}$, $\gamma_i$ and $m_i$ are the radial speed, Lorentz factor, and mass respectively, of the $i^{th}$ lepton and are defined in Appendix~\ref{sec:COMcalc}. Due to the small opening angle of our simulated fireball, the COM velocity (bulk velocity) can be written as --
\be
\vec{V}_{\rm COM} = V_{\rm COM} \hat{z}.
\ee
Using this bulk velocity, we can Lorentz transform the four--momenta of the particle to the COM frame. This transformation removes the radial--bulk motion component from the particle momentum, leaving behind the momentum only due to random motion. Let us denote the random velocity of the $i^{th}$ lepton (as measured in the COM frame) by $v^{\prime}_{\rm i/COM}$. The Lorentz factor associated with this random motion is --
\be
\gamma^{\prime}_{\rm i/COM} = \frac{1}{\sqrt{1 - \frac{v^{\prime 2}_{\rm i/COM}}{c^2}}}.
\ee
The average random kinetic energy or internal energy (denoted by $<KE^{\prime}>$) of the system can be computed as --
\be
<KE^{\prime}> = \frac{\sum_{i=0}^{\rm N_e} KE^{\prime}_i}{\rm N_{\rm e}} = {\sum_{i=0}^{\rm N_e}} \frac{\left(<\gamma^{\prime}_{\rm i/COM}> - 1\right) m c^2}{N_{\rm e}}, 
\ee
where $KE^{\prime}_i$ is the random kinetic energy of the $i^{th}$ lepton and $\rm N_{\rm e}$ denotes the total number of leptons present in the fireball. As the average kinetic energy and temperature are related by --
\be
<KE^{\prime}> \simeq 3 k_B T^{\prime},
\ee
the average temperature $T^{\prime}$ of the leptons can be computed as --
\be
T^{\prime} \simeq \frac{\sum_{i=0}^{\rm N_e} \left(<\gamma^{\prime}_{\rm i/COM}> - 1\right) m c^2}{3 k_B N_{\rm e}}, 
\ee
where $k_B$ denotes the Boltzmann's constant.

\section{Code Tests}\label{sec:sub_codetest}
\subsection{Temperature Evolution}\label{sec:sub_ctest_tempevol}
\begin{figure}
	\centering
	\includegraphics[width=1.0\linewidth]{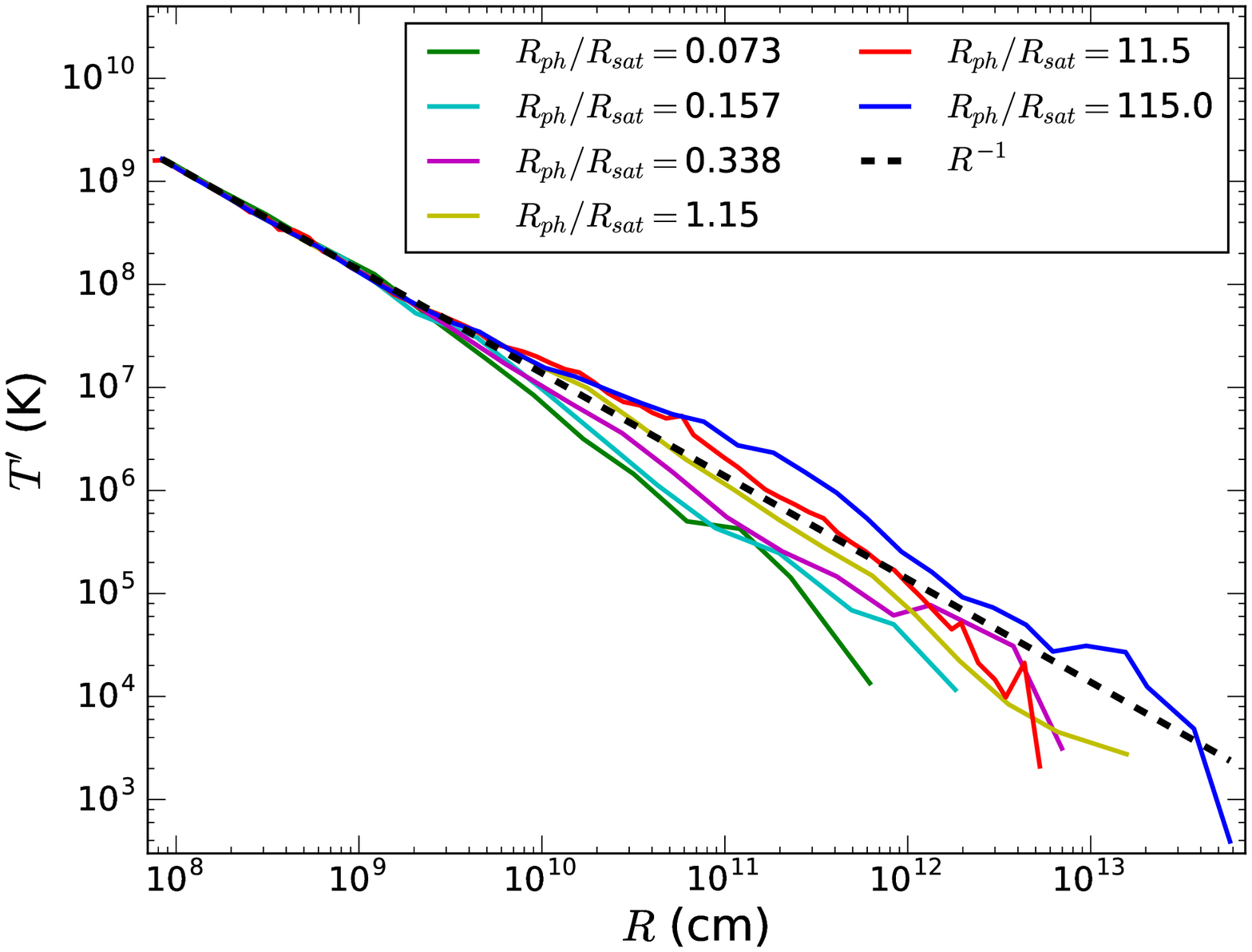}
	\caption{Evolution of the comoving temperature with the fireball radius. Each colored curve represents a fireball that begins evolution with a unique initial opacity. Also shown (by the dashed line) are the fireball model's theoretical predictions during the radiation--dominated acceleration phase.}
	\label{fig:nonradialtemp_onlye}
\end{figure}
\begin{figure}
	\centering
	\includegraphics[width=1.0\linewidth]{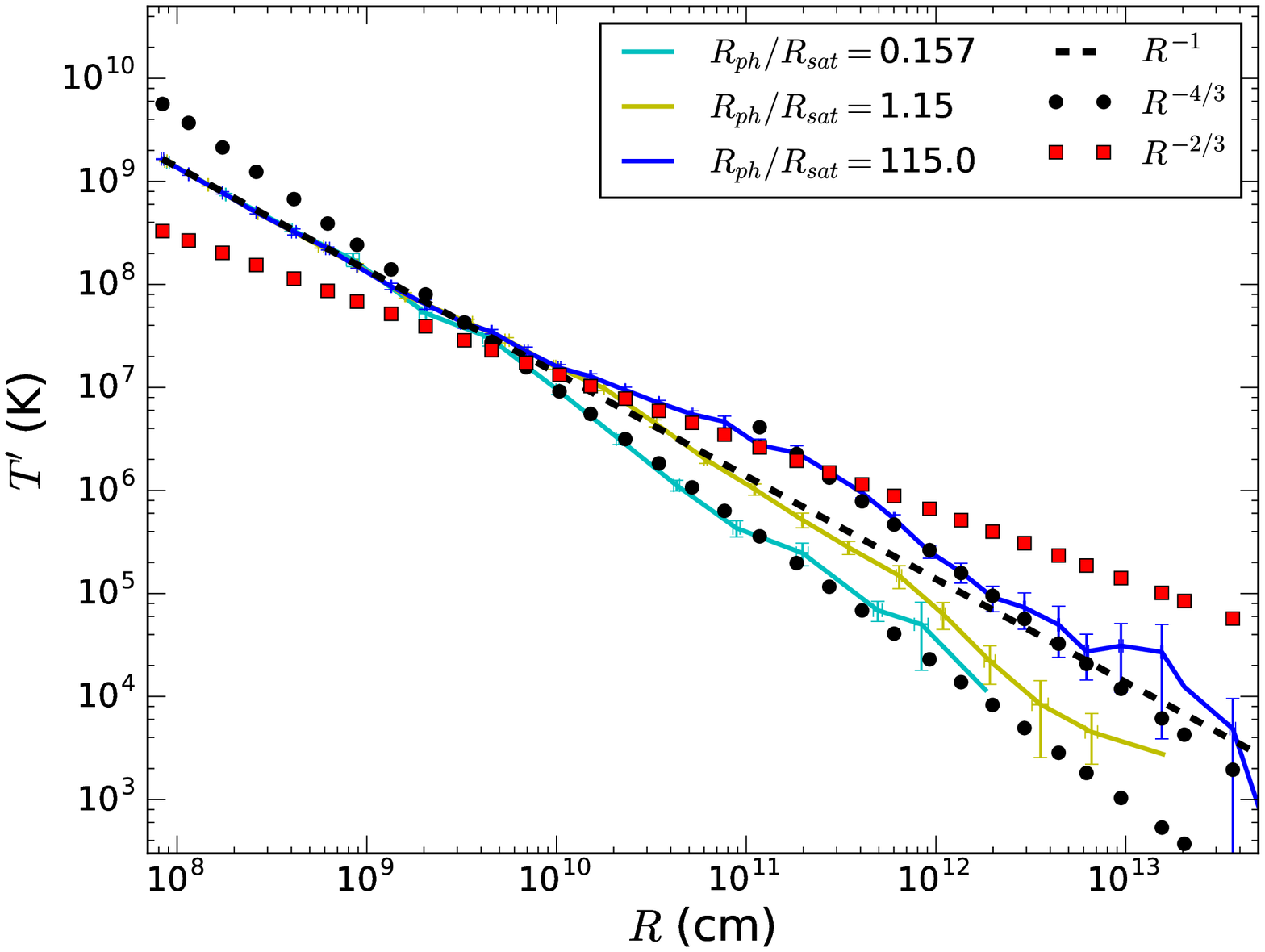}
	\caption{Comoving temperature evolution with radius for fireballs with $R_{\rm ph} / R_{\rm sat} = 0.157, 1.15~\&~115$. The black circles depict a line that decays proportionally to $R^{-4/3}$, the red squares depict a decay proportional to $R^{-2/3}$, and the black dashed line represents a curve evolving proportionally to $R^{-1}$. The deviations of the evolutionary trajectories from the theoretical predictions are discussed in \S~\ref{sec:sub_ctest_tempevol}. Also plotted are the error bars for the simulations computed from their standard deviation.}
	\label{fig:tempevol_special}
\end{figure}
This section explores the evolution of comoving temperature $T^{\prime}$ (as measured in the jet frame) with the radius of the fireball. Figure~\ref{fig:nonradialtemp_onlye} depicts the comoving temperature evolution of baryon--free fireballs, parameterized by their different initial opacities (as discussed earlier in \S~\ref{sec:resanddiscuss}, baryon--loaded fireball evolution is similar to the evolution of baryon--free fireballs, and hence the results of one hold true for the other). Fig.~\ref{fig:nonradialtemp_onlye} shows that during the radiation--dominated acceleration phase the comoving temperature of all fireballs grows proportionally to $R^{-1}$, thus verifying the fireball model's proportionality relations (see eq.~\ref{eq:fireball_acc}). The fireballs deviate from $T^{\prime} \propto R^{-1}$ path when the radiation dominance ends (as discussed in \S~\ref{sec:resanddiscuss}). However, from  Fig.~\ref{fig:nonradialtemp_onlye} we note that fireballs deviate differently depending upon their initial opacity.\\
Figure~\ref{fig:tempevol_special} investigates these deviations in greater detail. Similar to Fig.~\ref{fig:nonradialtemp_onlye}, Fig.~\ref{fig:tempevol_special} plots the comoving temperature with the fireball radius, but only for three fireballs (instead of six) characterized by $R_{\rm ph}/R_{\rm sat} = 0.157, 1.15~\&~115.0$. Only three fireballs are plotted to improve clarity and develop a better understanding of the comoving temperature deviations. For each fireball displayed in Fig.~\ref{fig:tempevol_special}, the transition from the radiation--dominated phase to the transient phase produces different evolutionary relationships between $T^{\prime}$ and $R$. If the fireball transitions from quadrant 1 to 3 via 2 (see Fig.~\ref{fig:Dyanmofbphasespace}, during the post--photospheric acceleration phase) the reduced optical depth causes radiation to decouple. As a result, the cyan curve (the least optically thick of the plotted fireballs) drops below the $T^{\prime} \propto R^{-1}$ dashed line, and begins to evolve along the black dotted circles representing $T^{\prime} \propto R^{-4/3}$ path.\\
The blue curve follows a different evolutionary path, a consequence of the blue fireball transitioning from quadrant 1 to 3 via quadrant 4 (see Fig.~\ref{fig:Dyanmofbphasespace}). The blue curve represents a highly opaque fireball that achieves matter domination before transparency, and as a result, undergoes the Thomson--dominated acceleration phase (see \S~\ref{sec:resanddiscuss}). This results in the average lepton energy exceeding the photon energy and the fireball accelerates (as well as cools) gradually such that $ T^{\prime} \propto R^{-2/3}$. During the matter--dominated phase, the blue curve in Fig.~\ref{fig:tempevol_special} jumps above the $T^{\prime} \propto R^{-1}$ dashed line, and follows the red squares. However, when the fireball eventually enters the optically thin regime (transitioning from quadrant 4 to 3) it begins following the black circles.\\
The yellow curve represents a fireball ($R_{\rm ph}/R_{\rm sat} = 1.15$) that is  comparatively optically thick than the cyan one. As a result, its initial deviation (from the radiation--dominated phase) is similar to blue curve than the cyan curve. Fig.~\ref{fig:Dyanmofbphasespace} shows that the yellow curve becomes matter--dominated before becoming optically thin, and transitions from quadrant 1 to 3 via 4. As a result, the yellow curve jumps over the $T^{\prime} \propto R^{-1}$ (dashed) line just like the blue curve, but only slightly. As the fireball represented by the yellow curve is not as optically thick as the one represented by the blue curve, the yellow curve becomes optically thin at a comparatively smaller radius. As a result, it drops below the $T^{\prime} \propto R^{-1}$ line earlier than the blue fireball, and begins evolving along the black circles. Thus, we find that the fireball's transient phases can explain the temperature evolution and its deviation from $T^{\prime} \propto R^{-1}$ proportionality relation.
\subsection{Shell Width Evolution}\label{sec:sub_ctest_width}
\begin{figure}
	\centering
	\includegraphics[width=1.0\linewidth]{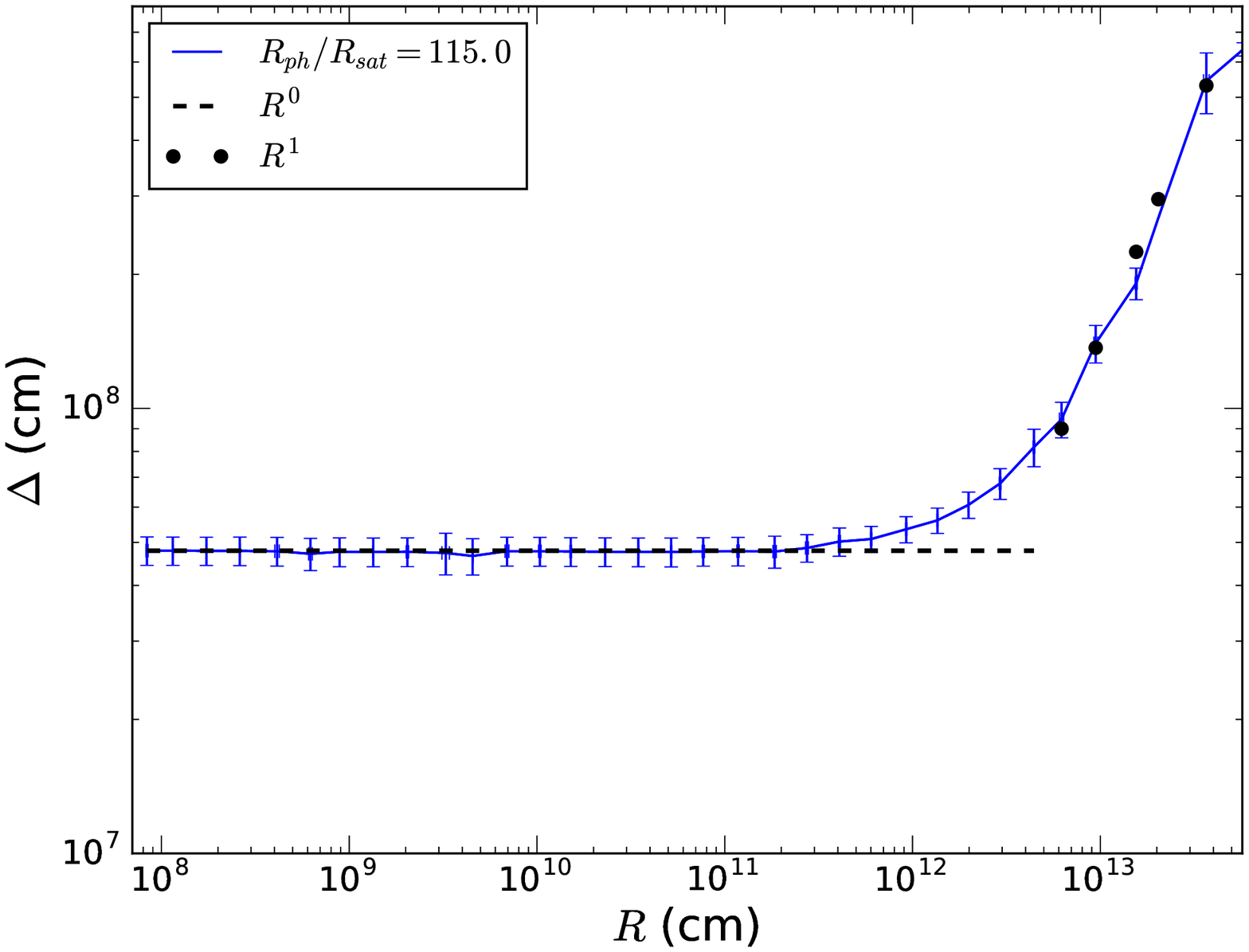}
	\caption{Evolution of the width of a fireball shell (for $R_{\rm ph}/R_{\rm sat} = 115$) as the fireball expands. As predicted by theory, the shell maintains a constant width during the acceleration phase.}
	\label{fig:width-vs-rad}
\end{figure}
\cite{MesLagunaRees93} found that the width of a fireball shell in the lab--frame remains constant during the radiation--dominated acceleration phase. The shell starts expanding proportionally to $R$ during the matter--dominated phase. Figure~\ref{fig:width-vs-rad} depicts the evolution of the width of the shell $\Delta$ in the lab frame and its evolution with the fireball radius, and it shows that the model is in agreement with the theory barring the transition phase.

\subsection{Restarted Simulations}\label{sec:sub_ctest_restart}
As shown in \S~\ref{sec:resanddiscuss}, the DynaMo code reproduces the proportionality relations of the fireball model (see eq.~\ref{eq:fireball_acc}). A powerful feature of the code is the ability to start (or restart) the simulations at a specific point along the simulation (positions and four--momenta of particles are all that is required to initialize the simulation). In this section we discuss the results obtained by restarting a completed simulation at a specific point along its evolution and compare them with the results of the original simulation.\\
The simulations we are considering in this section are specified by $R_{\rm ph} / R_{\rm sat} = 0.157$ (see Fig.~\ref{fig:grad-v-rrad-diff-sims}). The specific data point along the evolution which we use for the starting conditions for the \lq restarted simulation\rq~refers to the point when the bulk Lorentz factor of the original simulation equals 3.
\begin{figure}
	\centering
	\includegraphics[width=1.0\linewidth]{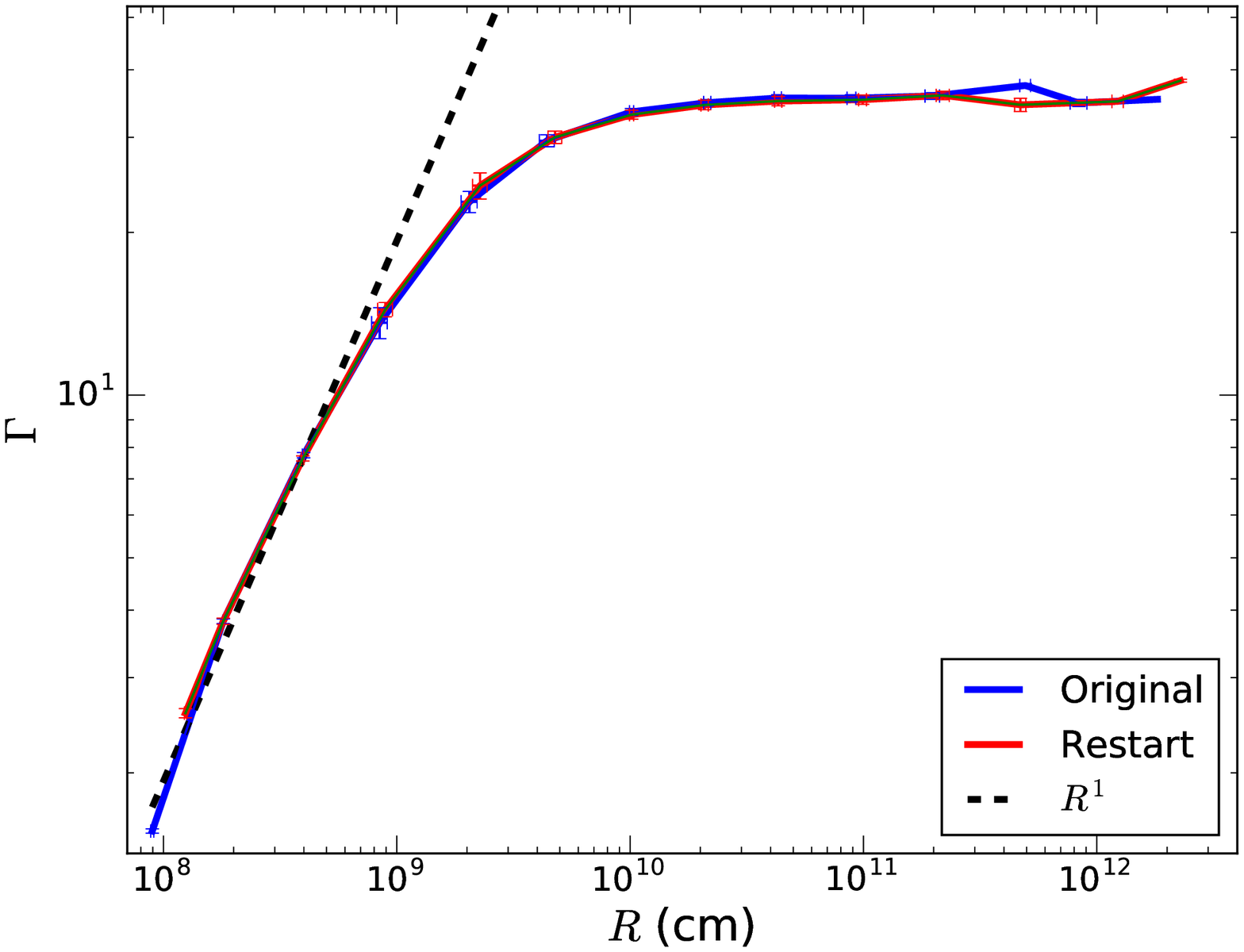}
	\caption{Evolution of Lorentz factor with radius for fireballs with $R_{\rm ph} / R_{\rm sat} = 0.157$. The blue curves represents the original simulation, which starts with no bulk motion. The red curve represents the simulation \lq restarted\rq~from the original simulation with initial conditions extracted at $\Gamma \sim 3$. Also plotted are the error bars for both the simulations computed from their standard deviation.}
	\label{fig:restart_grad-v-rrad-diff-sims}
\end{figure}
\begin{figure}
	\centering
	\includegraphics[width=1.0\linewidth]{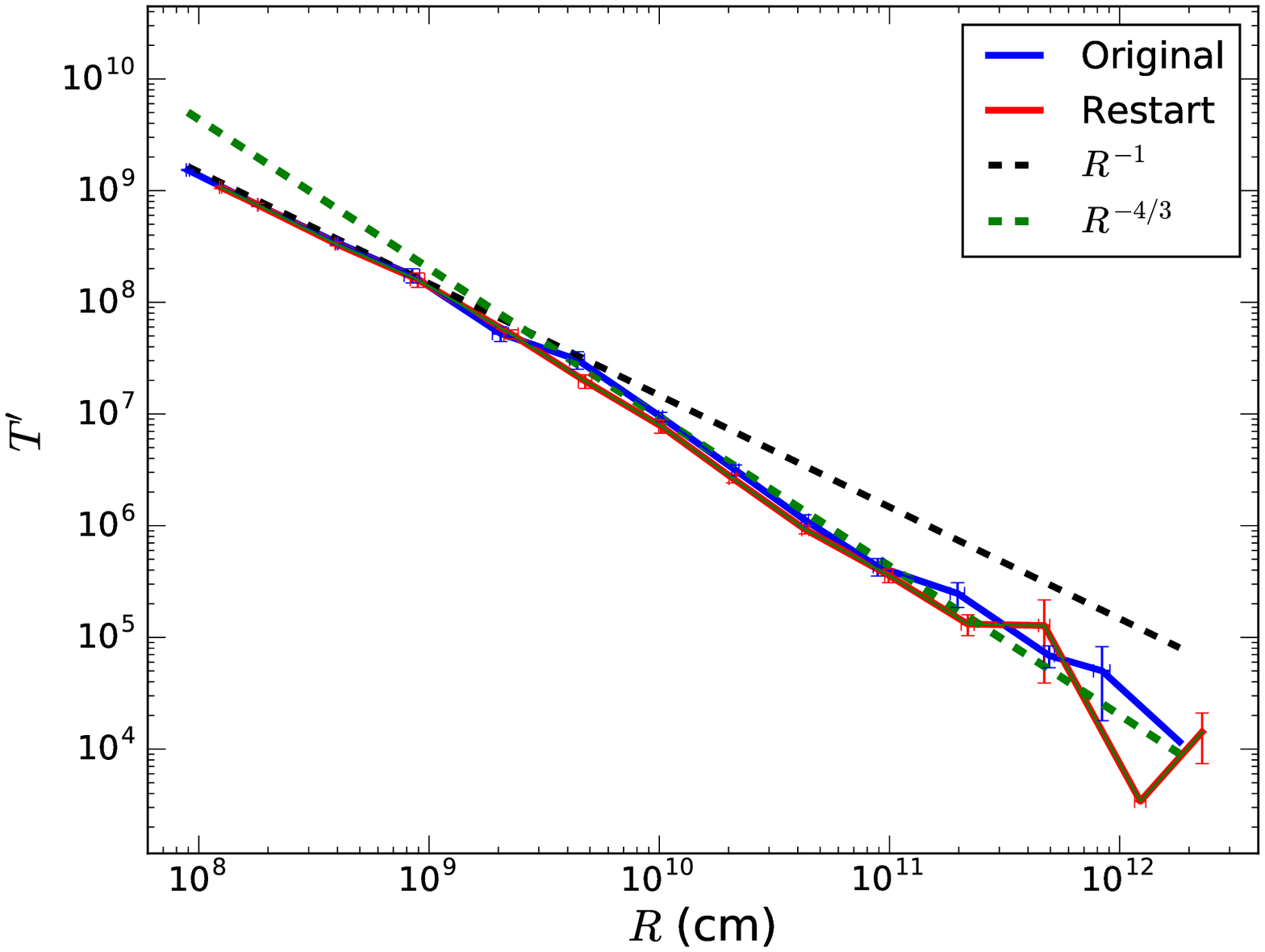}
	\caption{Evolution of comoving temperature with radius for fireballs with $R_{\rm ph} / R_{\rm sat} = 0.157$. The blue curves represents the original simulation which started with no bulk motion. The red curve is started with initial conditions extracted from the data of the original simulation (when $\Gamma \sim 3$). Also plotted are the error bars for the two simulations (computed from their standard deviation).}
	\label{fig:restart_tngrad-v-rrad-diff-sims}
\end{figure}
As shown in Figs.~\ref{fig:restart_grad-v-rrad-diff-sims} and~\ref{fig:restart_tngrad-v-rrad-diff-sims}, the two simulations evolve consistently as expected, and both simulations are in agreement with the proportionality relations of the fireball model.
\section{Calculation of the $R_{\rm ph}$ and $R_{\rm sat}$}\label{sec:calcRphRsat}
A relativistically expanding fireball achieves saturation when all its internal energy is converted into bulk motion, and as a result its bulk Lorentz factor achieves the maximum possible value. 
Using the fireball initialization parameters defined in \S~\ref{sec:jet_init}, the maximum Lorentz factor achieved at saturation in our simulation is $\eta \sim 112.9 $. 
In this section we calculate the photospheric radius $R_{\rm ph}$ in terms of the parameters defined in \S~\ref{sec:jet_init}. As the photospheric radius evolves differently during the acceleration and saturation phases of the fireball evolution, therefore we consider two cases: 1) the fireball attains the photospheric radius before saturation, i.e., $R_{\rm ph}<R_{\rm sat}$ and 2) when fireball achieves saturation before reaching the photospheric radius $R_{\rm ph} \geq R_{\rm sat} $. 
\subsection{Case I: $ R_{\rm ph}/R_{\rm sat}<1$}\label{sec:calcRphRsatI}
The first case considers the scenario that the photospheric radius is reached before the saturation radius, i.e., $R_{\rm ph}<R_{\rm sat}$. The saturation radius $R_{\rm sat}$ (the radius at which the fireball attains $\eta$) has already been obtained in eq.~\ref{eq:rsat} as --
\be
R_{\rm sat} = \eta \frac{R_0}{\Gamma_0} = \frac{E}{M c^2} \frac{R_0}{\Gamma_0}.
\ee
The calculation of the photospheric radius $R_{\rm ph}$ begins with the calculation of opacity of the evolving fireball. As stated in \cite{AbramNovikPaczy91}, the lab frame opacity $\tau$ is given by --
\be
\tau = \int_{R_{\rm ph}}^{\infty} n \sigma \left(1 - \beta \cos\theta \right) dR, \label{eq:opacity_rph_rsat}
\ee
where $n$ is the number density of scattering particles and the $\left(1 - \beta \cos\theta \right)$ term accounts for velocity dependence of opacity ($\beta$ is the bulk speed of the outflow normalized by the speed of light). Assuming $n R^2 = n_0 R_0^2 = $ constant (where $n_0$ is the initial particle density), and for $\cos{\theta} \sim 1$ (small angle approximation), eq.~\ref{eq:opacity_rph_rsat} can be written as --
\be
\tau = \int_{R_{\rm ph}}^{\infty} \frac{n_0 R_0^2 \sigma }{2 R^2 \Gamma_{\rm ph}^2} dR.\label{eq:opacity_rph_rsat+1}
\ee
If the fireball is within the acceleration regime (see eq.~\ref{eq:fireball_acc}), we can write --
\be
\frac{\Gamma_{\rm ph}}{R} = \frac{\Gamma_0}{R_0}\label{eq:gammaph_prop_rph}.
\ee
Combining eqs.~\ref{eq:opacity_rph_rsat+1} and~\ref{eq:gammaph_prop_rph}, the opacity equation becomes --
\be
\tau = \frac{n_0 R_0^2 \sigma}{2} \int_{R_{\rm ph}}^{\infty} \frac{1}{R^4}.
\ee
The photospheric radius $R_{\rm ph}$ is attained when $\tau \sim 1$, which implies --
\be
R_{\rm ph} = \left(\frac{n_0 R_0^4 \sigma}{6}\right)^{1/3}.
\ee
The ratio of the photospheric and the saturation radii can be written as --
\be
\frac{R_{\rm ph}}{R_{\rm sat}} = \left(\frac{n_0 R_0^4 \sigma}{6}\right)^{1/3}\frac{1}{\eta R_0}\label{eq:rphrsat_ratio<1}.
\ee
To vary the results of eq.~\ref{eq:rphrsat_ratio<1} we use $\sigma$ as an effective cross--section. The connection between the effective cross--sectional parameter $\sigma$ and the Thomson cross--section $\sigma_T$ is --
\be
\sigma_ \equiv \frac{\sigma_T \sigma_{\rm coeff} V_0}{c},\label{eq:ns_coeff}
\ee
where $V_0$ is the initial volume of the fireball shell and $\sigma_{\rm coeff}$ is the parameter we employ to scale the cross--section (and regulate the number of scatterings).
Re--writing eq.~\ref{eq:rphrsat_ratio<1} using eq.~\ref{eq:ns_coeff} --
\be
\frac{R_{\rm ph}}{R_{\rm sat}} = \left(\frac{n_0 R_0^4 \sigma}{6}\right)^{1/3}\frac{1}{\eta R_0} = \left(\frac{N R_0^4 \sigma_{\rm eff}}{6 c}\right)^{1/3}\frac{1}{\eta R_0}.
\ee
In our simulations we have kept the injection radius of, and the number of scatterers $N$ (and therefore the initial number density $n_0 V_0 = N$) in the fireball as fixed quantities. In order to explore fireballs with different opacity, we have to vary the quantity $\sigma$ by varying the parameter $\sigma_{\rm coeff}$ as they are proportional. The physical effect of increasing (decreasing) this parameter is to increase (decrease) the opacity (and as a consequence also the number of scatterings). As an example, for $\sigma_{\rm coeff}=10^5$ the ratio $R_{\rm ph}/R_{\rm sat} = 0.073$, and $R_{\rm ph}/R_{\rm sat} = 0.338$ results from $\sigma_{\rm coeff} = 10^7$ (see fig.~\ref{fig:grad-v-rrad-diff-sims}). \\

\subsection{Case II: $R_{\rm ph}/R_{\rm sat}\geq1$}
We now explore the case when the $R_{\rm ph}\geq R_{\rm sat}$. At the saturation radius, as all the internal energy has been converted into bulk kinetic energy, the acceleration of the fireball ceases and the fireball coasts beyond that radius at $\eta$. If the saturation radius is smaller than the photospheric radius, this implies radiation is still trapped within the fireball. As the fireball is no longer in the acceleration regime (see eq.~\ref{eq:fireball_acc}), $\Gamma_{\rm ph} = \eta$, and the fireball opacity evolves differently from the case $R_{\rm ph}<R_{\rm sat}$. Using eq.~\ref{eq:opacity_rph_rsat}, we can write --
\be
\tau = 1 = \int_{R_{ph}}^{\infty} \frac{n_0 R_0^2 \sigma}{2 R^2 \eta} dR.
\ee
The photospheric radius obtained by integration can be expressed as --
\be
R_{\rm ph} = \frac{n_0 \sigma R_0^2}{2 \eta^2}.
\ee
Therefore, the ratio of the photospheric to the saturation radii is --
\be
\frac{R_{\rm ph}}{R_{\rm sat}} = \frac{n_0 \sigma R_0}{2 \eta^3}\label{eq:rphrsat_ratio>1}. 
\ee
Using eq.~\ref{eq:ns_coeff}, we can rewrite eq.~\ref{eq:rphrsat_ratio>1} as --
\be
\frac{R_{\rm ph}}{R_{\rm sat}} = \frac{n_0 \sigma R_0}{2 \eta^3} = \frac{N \sigma_{\rm eff} V_0 R_0}{2 c V_0\eta^3} = \frac{N \sigma_{\rm eff} R_0}{2 c \eta^3}.
\ee
As discussed in Appendix~\ref{sec:calcRphRsatI}, by tuning the parameter $\sigma_{\rm coeff}$ we can regulate the fireball opacity. For example, by using $\sigma_{\rm coeff} = 10^8$ we produce the ratio $R_{\rm ph}/R_{\rm sat} = 1.15$. Similarly, by increasing $\sigma_{\rm coeff}$ to $10^{10}$ we obtain an increased ratio $R_{\rm ph}/R_{\rm sat} = 115$.

\bsp	
\label{lastpage}
\end{document}